\documentclass[hyper,11pt,letter]{article}
\pdfoutput=1
\usepackage{jheppub} 

\usepackage[T1]{fontenc} 
\usepackage{graphicx,amsmath,amssymb,multirow,array,bm,bbm,mathrsfs}
\usepackage[mathscr]{eucal}

\usepackage{epstopdf}
\usepackage{array,tikz-cd}
\usepackage{subfigure} 

\newcommand*{\boxcolor}{black}
\makeatletter
\renewcommand{\boxed}[1]{\textcolor{\boxcolor}{%
\tikz[baseline={([yshift=-1ex]current bounding box.center)}] \node [rectangle, minimum width=1ex,rounded corners,draw] {\normalcolor\m@th$\displaystyle#1$};}}
 \makeatother

\newcommand{\epsS}{{\cal E}}

\newcommand{\PP}{P_{_\text{P}}}
\newcommand{\PR}{P_{_\text{R}}}
\newcommand{\PH}{P_{_\text{H}}}

\makeatletter
\newcommand{\doublewidetilde}[1]{{%
  \mathpalette\double@widetilde{#1}%
}}
\newcommand{\double@widetilde}[2]{%
  \sbox\z@{$\m@th#1\widetilde{#2}$}%
  \ht\z@=.9\ht\z@
  \widetilde{\box\z@}%
}
\makeatother

\title{Reparametrization modes, shadow operators, and quantum chaos in higher-dimensional CFTs}

\author[a,b]{Felix M.\ Haehl}
\author[b]{, Wyatt Reeves}
\author[b]{, Moshe Rozali}

\affiliation[\,a]{School of Natural Sciences, Institute for Advanced Study,\\
Einstein Drive, Princeton, NJ 08540, USA.}
\affiliation[\,b]{Department of Physics and Astronomy, University of British Columbia,\\
6224 Agricultural Road, Vancouver, BC V6T 1Z1, Canada.}

\emailAdd{f.m.haehl@gmail.com}
\emailAdd{wreeves@phas.ubc.ca}
\emailAdd{rozali@phas.ubc.ca}

\abstract{We study two novel approaches to efficiently encoding universal constraints imposed by conformal symmetry,  and describe applications to quantum chaos in higher dimensional CFTs. The first approach consists of a reformulation of the shadow operator formalism and kinematic space techniques. We observe that the shadow operator associated with the stress tensor (or other conserved currents) can be written as the descendant of a field $\epsS$ with negative dimension. Computations of stress tensor contributions to conformal blocks can be systematically organized in terms of the ``soft mode'' $\epsS$, turning them into a simple diagrammatic perturbation theory at large central charge. 

Our second (equivalent) approach concerns a theory of reparametrization modes, generalizing previous studies in the context of the Schwarzian theory and two-dimensional CFTs. Due to the conformal anomaly in even dimensions, gauge modes of the conformal group acquire an action and are shown to exhibit the same dynamics as the soft mode $\epsS$ that encodes the physics of the stress tensor shadow. We discuss the calculation of the conformal partial waves or the conformal blocks using our effective field theory. The separation of conformal blocks from shadow blocks is related to gauging of certain symmetries in our effective field theory of the soft mode.

These connections explain and generalize various relations between conformal blocks, shadow operators, kinematic space, and reparametrization modes.  As an application we study thermal physics in higher dimensions and argue that the theory of reparametrization modes captures the physics of quantum chaos in Rindler space. This is also supported by the observation of the pole skipping phenomenon in the conformal energy-energy two-point function on Rindler space.}

\begin{document} 
	
\maketitle
\flushbottom

\section{Introduction}

Chaos in dynamical systems is a crucial phenomenon explaining thermalization in many-body systems. There is a large number of manifestations of quantum and classical chaos, which can be roughly separated into early- and late-time characteristics.  At long times the asymptotic approach to equilibrium is described in the framework of an effective field theory of the long-lived modes, hydrodynamics. 

In this paper we discuss some early-time manifestations of quantum chaos. The relevant phenomena are characterized by a so-called scrambling time of order the logarithm of the entropy, and can be quantified by the exponential growth of certain out-of-time-order correlation functions (OTOCs) \cite{Larkin:1969aa,Shenker:2013pqa,Shenker:2013yza,Leichenauer:2014nxa,Maldacena:2015waa,Kitaev:2015aa,Hayden:2007cs,Sekino:2008he,Lashkari:2011yi}. The associated growth rate is dubbed the Lypaunov exponent, which was shown to be bounded from above using only general principles of quantum field theory \cite{Maldacena:2015waa}.

Thus, maximally chaotic theories are special. For example, theories dual to Einstein gravity are maximally chaotic, which leads to the intuition that a detailed understanding of maximal, or near-maximal chaos can perhaps be utilized to explain the emergence of a holographic dual to many-body strongly interacting systems.

Another, perhaps related, special property of maximally chaotic theories is the existence of an effective field theory of a single mode, sometimes referred to as the `soft mode', the `scramblon', or the `reparametrization mode'. This is somewhat surprising since we are away from the long time limit justifying the use of the hydrodynamic effective description. Rather, the small parameter controlling the effective theory of the soft mode is the inverse of large number of degrees of freedom, which also controls the thermodynamic limit.

The first example of such an effective description is the Schwarzian theory, describing the low energy limit of the SYK model \cite{Kitaev:2015aa,Maldacena:2016hyu}, or the boundary dynamics of Jackiw-Teitelboim gravity \cite{Almheiri:2014cka,Maldacena:2016upp,Jensen:2016pah}. Based on this example \cite{Blake:2017ris} suggested a general effective field theory description of maximal chaos in theories with conserved energy, by extrapolating hydrodynamics to the relevant short time regime. This description was named ``quantum  hydrodynamics'' due to its formal similarity with classical hydrodynamics. 

Given the success of the Schwarzian theory, reparametrization modes have also been used in two-dimensional CFTs for various applications such as conformal blocks \cite{Cotler:2018zff,Jensen:2019cmr} and OTOCs (in thermal states) \cite{Turiaci:2016cvo,Haehl:2018izb}. While all these applications are clearly related to the physics of stress tensor conservation, a detailed understanding of this connection is missing. In this work we aim to explore the physics of reparametrization modes from the point of view of conformal representation theory, thus elucidating the connection between numerous different calculations in the literature. In order to uncover the general principles, it is convenient to work in an arbitrary number of (even) dimensions.

\paragraph{Summary.}
In this paper we aim to provide a detailed understanding of the underlying physics of the reparametrization mode in CFTs. We clarify the connection with a number of other approaches towards to the computation and organization of conformal correlators. We argue that the reparametrization mode can be understood as a ``longitudinal'' (pure gauge) contribution to the {\it shadow operator associated with the stress tensor}. The shadow of the stress tensor is a formal operator of dimension $0$, which allows for a convenient projection of correlation functions onto their contribution coming from exchanges of the stress tensor and its descendants alone \cite{Dolan:2011dv,SimmonsDuffin:2012uy}. This connection explains why the reparametrization mode is useful for the computation of global or Virasoro conformal blocks. 

We discuss two approaches to defining the soft mode in higher dimensional CFTs, and describe its quadratic action. The first relates the soft mode to the shadow of the energy momentum tensor, and the second defines it directly in terms of reparametrizations. The two approaches are shown to be equivalent, and generalize the Schwarzian in one dimension, and the Alekseev-Shatashvili action \cite{Alekseev:1988ce,Cotler:2018zff} in two dimensions.

We then describe the coupling of the soft modes to external operators. In any CFT, the projection (fusion) of two primary probe operators into a stress tensor is described by the operator product expansion (OPE). The basic three-point fusion into a stress tensor is universally captured by the bilinear {\it stress tensor OPE blocks}. These operators, depending on two insertion points, have previously been studied in the context of {\it kinematic space}, where they appear naturally \cite{Czech:2016xec,deBoer:2016pqk}. We argue that they can be thought of as the ``vertices'' coupling reparametrization modes to external operators. In low dimensional examples, these vertices have previously been understood as reparametrized two-point functions. 

We use our machinery to describe conformal blocks in arbitrary even dimensions. The basic ingredients we develop calculate simply the conformal partial waves, a certain combination of the stress energy block and it shadow. We argue that to isolate the physical block from its shadow, certain symmetries we call ``conformal redundancies'' need to be gauged. We demonstrate that this prescription give the correct answer in two dimensions, and make comments on those symmetries in higher dimensional CFTs.

Finally, we use our novel understanding of reparametrization modes and conformal kinematics to draw some conclusions about thermal physics in higher dimensions. Both a pole skipping analysis of energy-energy correlation functions, and  the propagation of reparametrization modes between bilocal OPE block operators, lead us to a derivation of the Lyapunov exponent and butterfly velocity describing theories which exhibit maximal chaos in Rindler space. As in two dimensions, maximal chaos is, of course, not universal. Our analysis only captures the stress tensor contribution to the out of time order correlator. This establishes a connection between our consideration of conformal representation theory, and the ``quantum hydrodynamic'' description of maximally chaotic theories.

\paragraph{Outline.}
In \S\ref{sec:shadows} we study conformal kinematics as encoded in global conformal blocks and partial waves. We provide a convenient reformulation of the shadow operator formalism for the stress tensor conformal block in terms of a vector mode with negative dimension, and use it to calculate the associated conformal partial wave in arbitrary even dimensions. We show in \S\ref{sec:reparametrizations} that this negative dimension ``shadow mode'' can be understood as a reparametrization mode very similar to the one that occurs in the Schwarzian theory or two-dimensional CFTs. We derive a quadratic action for this mode from the conformal anomaly. We study the theory of reparametrizations in more detail for the case of two-dimensional theories in \S\ref{sec:2d}, with particular interest in the conformal map to the thermal state. In \S\ref{sec:thermal} we generalize this analysis to higher dimensions by mapping the Rindler wedge to a hyperbolic spacetime, thus creating a state with non-zero temperature. In \S\ref{sec:EEcorr} we demonstrate pole skipping of the energy-energy two-point function in CFTs in an arbitrary number of dimensions. The pole skipping location allows us to read off a maximal Lyapunov exponent $\lambda_L = \frac{2\pi}{\beta}$ and the Rindler space butterfly velocity $v_B = \frac{1}{d-1}$. We finish with comments and questions in \S\ref{sec:conclusion}.

\vspace{10pt}
\section{Global conformal blocks and the shadow of the stress tensor}
\label{sec:shadows}

We begin our discussion with an exploration of global conformal blocks for CFTs in an arbitrary number of dimensions. Specifically, in this section we review and extend the kinematic space perspective on OPE blocks as bilocal operators. We offer a novel point of view on OPE blocks and conformal partial waves based on a reformulation of the shadow operator formalism. We will see that the shadow operators associated with conserved currents have special properties, which make them closely related to the reparametrization mode, i.e., the basic ingredient of the EFT of quantum chaos in the two-dimensional case \cite{Haehl:2018izb,Cotler:2018zff}. In the process we identify the soft mode in higher-dimensional CFTs and its coupling to matter (which is given by the OPE blocks). For the remainder of this section, we will work in Euclidean flat spacetime. The map to thermal physics in hyperbolic space and the relation with quantum chaos will be explored in subsequent sections.

\vspace{10pt}
\subsection{Review of kinematic space and OPE blocks}


We will be interested in the operator product expansion (OPE) of nearby operators. In order to efficiently organize the kinematic constraints that describe the universal aspects of the OPE in CFTs, it is useful to introduce the space on which pairs of operators live.

\paragraph{Kinematic space.}
One definition of the kinematic space ${\cal M}_\lozenge^{(d)}$ of $d$-dimensional CFTs is as the space of pairs of either timelike or spacelike separated points. The former naturally define causal diamonds with a spherical base region, so this space is clearly the same as the space of causal diamonds. A moment's thought reveals that this is in fact also isomorphic to the space of {\it spacelike} separated points, which define spacelike hyperbolas in one-to-one correspondence with causal diamonds.\footnote{  This is obvious in embedding space. See Appendix A.2 of \cite{deBoer:2016pqk} for details.} More invariantly, we can define kinematic space as the coset
\begin{equation}
\label{eq:kinspace}
  {\cal M}_\lozenge^{(d)} \equiv \frac{SO(2,d)}{SO(1,d-1) \times SO(1,1)} \,.
\end{equation}
Here, $SO(2,d)$ is the conformal group in $d$ spacetime dimensions. We mod out the group which leaves the objects in consideration invariant. For instance, in order to understand ${\cal M}_\lozenge^{(d)}$ as the space of codimension-2 spheres (i.e., causal diamonds), the coset should be understood as follows: The stabilizer of a given sphere consists of the group $SO(1,d-1)$ of rotations and special conformal transformations leaving the sphere on a fixed time slice invariant, as well as a group $SO(1,1)$ that corresponds to evolving the time slice along the conformal Killing flow inside the causal diamond. The resulting space is $2d$-dimensional with half of the directions being spacelike and the other half being timelike. For instance, in $d=2$, the above coset factors into two copies of de Sitter space $dS_2 \equiv SO(2,1)/ SO(1,1)$. We refer the reader to \cite{Czech:2016xec,deBoer:2016pqk} for more extensive studies of this construction and the associated geometry. 

For most of this paper we will be working in Euclidean signature. In that case, the picture of kinematic space as the space of spacelike separated points is most natural, as it can easily be analytically continued. As a coset, we define the kinematic space of a Euclidean CFT as
\begin{equation}
\label{eq:kinE}
  {\cal M}_{\lozenge,E}^{(d)} \equiv  \frac{SO(1,d+1)}{SO(1,d-1) \times SO(2)} \,,
\end{equation}
with the stabilizer group corresponding to the subgroup leaving pairs of (Euclidean) points invariant under rotations, special conformal transformations, and the modular flow whose only two fixed points are the two points under consideration.\footnote{ The modular flow leaving two points $(x^\mu,y^\mu)$ invariant is generated by the conformal Killing vector 
\begin{equation*}
   K^\mu(x,y;\xi) = -\frac{2\pi}{(y-x)^2} \left[ (y-\xi)^2 (x^\mu - \xi^\mu) - (x-\xi)^2 (y^\mu - \xi^\mu ) \right] \,.
\end{equation*}
In Lorentzian or Euclidean signature, the conformal Killing vector looks identical, but the inner product is taken with either the Lorentzian or Euclidean metric.}
Most objects that we discuss in this paper naturally live on the space \eqref{eq:kinE}.

\paragraph{OPE blocks as kinematic space operators.} 
Consider now a Euclidean CFT. An immediate question about the space of pairs of points $(x^\mu,y^\mu)$, is whether there exist natural operators on this space, i.e., bilocal operators in the CFT. This is indeed the case, and we get an important hint from the operator product expansion (OPE) of two operators $V,W$. The latter can be written as a sum over conformal families of primary operators and their descendants, which $V$ and $W$ have non-trivial three-point overlap with. Schematically, we write as $x^\mu \rightarrow y^\mu$:
\begin{equation}
\label{eq:OPE}
   V(x) W(y) = \frac{1}{|x-y|^{\Delta_V + \Delta_W }}  \sum_{{\cal O}} C_{VW {\cal O}} \; |x-y|^{ \Delta} \, \big[ {\cal O}(y) + \text{conformal descendants} \big]\,.
   \end{equation}
In the following we will restrict to pairwise equal operators, $V=W$, for simplicity. Since we will eventually be interested in the OPE block associated with stress tensor exchanges, only identical external operators give a non-trivial result. We will refer to the sum over descendants as the {\it OPE block} $\mathfrak{B}_{{\cal O}}(x,y) $:
\begin{equation}
\label{eq:OPEblocks}
   V(x) V(y) \equiv \frac{1}{|x-y|^{2\Delta_V }}  \sum_{{\cal O}} C_{VV {\cal O}} \;   \mathfrak{B}_{{\cal O}}(x,y) \,.
\end{equation}
The OPE is concerned with the limit $x^\mu \rightarrow y^\mu$. However, in abuse of language we will often refer to OPE blocks even if the two points $(x^\mu,y^\mu)$ are not near each other.
In fact, we will make more precise the fact that it is sensible and useful to study bilocal OPE blocks in the form of an integral over the associated conformal primary. We define:\footnote{  Eq.\ \eqref{eq:smearRep} is written with Euclidean signature in mind, where the integration region is all of $\mathbb{R}^d$ (see Appendix \ref{sec:conventions} for details about this notation). A natural Lorentzian generalization, appropriate for studying the OPE with $x^\mu$ and $y^\mu$ timelike separated, involves integration over the causal diamond defined by $x^\mu$ and $y^\mu$. However, in order to avoid subtleties regarding convergence and making the expression well-defined, we leave the interesting Lorentzian case for future work.}
\begin{equation}
\label{eq:smearRep}
   \mathfrak{B}_{{\cal O}}(x,y) \equiv \frac{k_{d-\Delta,\ell}}{\pi^{d/2} C_{\cal O}} \int d^d\xi \; I_{VV{\cal O}}^{\mu_1 \cdots \mu_{\ell}} (x,y;\xi) \; {\cal O}_{\mu_1 \cdots \mu_{\ell}}(\xi) \,,
\end{equation}
where $C_{{\cal O}}$ is the coefficient in the two-point function of ${\cal O}$, $\Delta$ is its operator dimension, and
\begin{equation}
\label{eq:kNorm}
k_{\Delta,\ell} = \frac{\Gamma(\Delta-1)}{\Gamma(\Delta+\ell-1)} \frac{\Gamma(d-\Delta+\ell)}{\Gamma(\Delta-\frac{d}{2})} \,
\end{equation}
is a normalization factor.\footnote{  Note in particular the stress tensor case: $k_{d,2} = [d(d-1)\Gamma(d/2)]^{-1}$ and $k_{0,2}= -(-)^{d/2}\Gamma(d+2)\Gamma(\frac{d}{2}+1)$.} The constant in \eqref{eq:smearRep} is chosen to give a convenient normalization (c.f., \cite{Czech:2016xec}).
The kernel is determined by conformal symmetry, and can be thought of as a normalized three-point function between $VV$, and an auxiliary operator $\widetilde{{\cal O}}$ with dimension $\widetilde{\Delta} = d-\Delta$: 
\begin{equation}
\label{eq:tt1}
  I_{VV{\cal O}}^{\mu_1 \cdots \mu_{\ell}} (x,y;\xi) = \frac{1}{C_{VV{\cal O}}} \frac{\langle V(x) V(y) \widetilde{{\cal O}}^{\mu_1 \cdots \mu_{\ell}}(\xi) \rangle}{\langle V(x) V(y) \rangle}  \,,
\end{equation}
which is independent of the operator $V$ and OPE coefficient $C_{VV{\cal O}}$. The operator $\widetilde{{\cal O}}$ is known as the {\it shadow} of ${\cal O}$. 
One of the insights of \cite{Czech:2016xec,deBoer:2016pqk} was the realization that the sum over descendants in the OPE \eqref{eq:OPEblocks} is in fact equivalent to the integral \eqref{eq:smearRep}. We will derive this below using related methods.

Despite our preference for the Euclidean setup, the smeared representation of the OPE block still has to be enjoyed with care (similarly for the Lorentzian OPE with spacelike separated points). Since we take the integral in \eqref{eq:smearRep} to cover all of Euclidean spacetime, we are making a subtle mistake: we include contributions for which the $V \times V$ OPE in $\langle V(x) V(y) \widetilde{{\cal O}}(\xi) \rangle$ appearing in the kernel \eqref{eq:tt1} is not valid! For instance, in radial quantization, a more careful analysis would require $|\xi| > |x|,|y|$. It was understood in \cite{SimmonsDuffin:2012uy} that the effect of this carelessness is that the OPE block computed as in \eqref{eq:smearRep} will end up being contaminated by certain unphysical {\it shadow contributions} that need to be projected out. These shadow contributions are almost as desired (they have the same conformal properties as the actual physical OPE block, including the same eigenvalue under the conformal Casimir), if it weren't for unphysical short distance behavior. Instead of doing the conformal integrals more carefully, we will therefore have to deal with the task of projecting out shadow contributions at the end of various computations. We will discuss this projection and its physical interpretation in more detail in the following.

\vspace{10pt}
\subsection{Conformal blocks, shadow blocks, and partial waves}

The integral representation of the OPE blocks can be used to decompose four-point functions into more primitive building blocks. Consider the Euclidean four-point function of two pairs of operators, $V$ and $W$. Using the OPE \eqref{eq:OPE} for $VV$ and for $WW$, the four-point function decomposes into {\it conformal blocks} $G_{\Delta}^{(\ell)}$ associated with the exchange of different primary operators ${\cal O}$ and their descendants:
\begin{equation}
\langle V(x_1)V(x_2)W(x_3)W(x_4)\rangle =  \frac{1}{x_{12}^{\Delta_V} x_{34}^{\Delta_W}}\sum_{{\cal O}} C_{VV{\cal O}} \, C_{WW{\cal O}} \; G_{\Delta}^{(\ell)}(u,v) \,,
\end{equation}
where $(\Delta,\ell)$ are the dimension and spin of the internal operator ${\cal O}$ and $x_{ij} \equiv x_i - x_j$. In the following we will drop the $i$ subscript.
The conformal block is a purely kinematical object and only depends on the conformally invariant cross ratios 
\begin{equation}
\label{eq:CRs}
  u = \frac{x_{12}^2 x_{34}^2}{x_{13}^2 x_{24}^2}  \,,\qquad v = \frac{x_{14}^2 x_{23}^2}{x_{13}^2 x_{24}^2} \,.
\end{equation}
In terms of OPE blocks, we can write the conformal block associated with exchanges of global conformal descendants as a two-point function of bilinears:
\begin{equation}
\label{eq:GfromKin}
   G_\Delta^{(\ell)}(u,v) =  \langle \mathfrak{B}_{{\cal O}}(x_1,x_2) \, \mathfrak{B}_{{\cal O}}(x_3,x_4) \rangle_\text{phys.} 
\end{equation}
where the subscript ``phys.'' indicates that we need to project out the shadow contributions mentioned at the end of the previous section (see below for details).
Apart from being manifestly conformally invariant, the conformal blocks have other defining features. They are eigenfunctions of the $SO(2,d)$ conformal Casimir operator  ${\cal C}_2 = L_{AB} L^{AB}$ with eigenvalue $-C_{\Delta,\ell} = \Delta(d-\Delta) - \ell(\ell+d-2)$. Below we will be interested in the block associated with the stress tensor, where $C_{d,2} = 2d$. In terms of the cross ratios $(u,v)$ the Casimir reads
\begin{equation}
\label{eq:Casimir}
\begin{split}
  \frac{1}{2} \, {\cal C}^2 &= \left(u(1+v)-(1-v)^2\right) \partial_v \, v \, \partial_v  -(1-u+v)u \, \partial_u \,u \, \partial_u  +2(1+u-v) uv \, \partial_u \partial_v  + d \, u \, \partial_u \,.
 \end{split}
\end{equation}
Note that the shadow of a primary ${\cal O}$ has a conformal block associated with it, which shares the same eigenvalue $C_{\widetilde{\Delta},\ell}=C_{\Delta,\ell}$. We will refer to this object as the {\it shadow block} associated with the auxiliary operator $\widetilde{{\cal O}}$, and denote it as $ G_{d-\Delta}^{(\ell)}$.

\paragraph{Conformal partial waves.} 
More explicitly, we can write an integral representation of the shadow operator associated with a symmetric-traceless operator ${\cal O}$ \cite{Dolan:2011dv,SimmonsDuffin:2012uy}:
\begin{equation}
\label{eq:QOhigherd}
   \widetilde{{\cal O}}^{\mu_1 \cdots \mu_\ell}(x) \equiv \frac{k_{\Delta,\ell} }{\pi^{d/2}}
    \int d^d y \; \frac{ \prod_{i=1}^\ell \left( \delta^{\mu_i\nu_i} (x-y)^2 - 2 (x-y)^{\mu_i}(x-y)^{\nu_i}\right) }{\left( (x-y)^2 \right)^{d-\Delta+\ell} }\;  {\cal O}_{\nu_1\cdots \nu_\ell} (y)
\end{equation}
where we take the normalization from \cite{Dolan:2011dv} such that the operation of forming the shadow squares to the identity ($\doublewidetilde{{\cal O}} = {\cal O}$).

The shadow block is simply the conformal block associated with the shadow operator. Both the shadow operator and its block are usually not physical. They present formal tools that are useful for computations. In order to get physical results, one needs to make sure to project out shadow contributions at the end of calculations.

For our purposes it will often be convenient to consider a particular linear combination of block and shadow block. We will refer to this  as the (normalized) {\it conformal partial wave (CPW)} and define it as
\begin{equation}
\label{eq:CPWdef}
  f_\Delta^{(\ell)}(u,v) \equiv
   \frac{1}{k_{d-\Delta,\ell}}\frac{\Gamma(\frac{\Delta+\ell}{2})^2}{\Gamma(\frac{d-\Delta+\ell}{2})^2} \; G_\Delta^{(\ell)}(u,v) + \frac{1}{k_{\Delta,\ell}}\frac{\Gamma(\frac{d-\Delta+\ell}{2})^2}{\Gamma(\frac{\Delta+\ell}{2})^2}\; G_{d-\Delta}^{(\ell)}(u,v) \,.
\end{equation}
Both terms in $f_\Delta^{(\ell)}$ are conformally invariant and eigenfunctions of the Casimir with eigenvalue $C_{\Delta,\ell}$. They are distinguished by their short distance behavior: 
\begin{equation}
\label{eq:shortdist}
   G_\Delta^{(\ell)} \;\stackrel{u \rightarrow 0 , \; v \rightarrow 1}{\sim} \;  u^{\frac{\Delta-\ell}{2}} (1-v)^\ell + \ldots \,\qquad \quad
      G_{d-\Delta}^{(\ell)} \;\stackrel{u \rightarrow 0 , \; v \rightarrow 1}{\sim} \;  u^{\frac{d-\Delta-\ell}{2}} (1-v)^\ell + \ldots 
\end{equation}
The advantage of the conformal partial wave combination is that it is single-valued and arises naturally in computations in the shadow formalism. Indeed, the two-point function $\langle \mathfrak{B}_{{\cal O}}(x_1,x_2) \, \mathfrak{B}_{{\cal O}}(x_3,x_4) \rangle $ before projection is proportional to the CPW $f_\Delta^{(\ell)}$. The projection onto the physical block in \eqref{eq:GfromKin} corresponds to dropping the shadow block which has the wrong short distance behavior.

To summarize, conformal blocks are two-point functions of bilinear blocks $\mathfrak{B}_{\cal O}(x,y)$. These are global blocks in the sense that they correspond to a single stress tensor exchange. We discuss corrections to this leading answer later on. If we work in Euclidean signature, a monodromy projection onto the physical block is always necessary. At the level of the conformal blocks, this can be implemented as explained in \cite{SimmonsDuffin:2012uy} by simply projecting onto the conformally invariant part of $f_\Delta^{(\ell)}$ with the correct monodromy.

\paragraph{The stress tensor block.}
In this paper we mainly study OPE blocks of the stress tensor (see however \S\ref{sec:higherspin} for more general conserved currents). In this case, the shadow of the stress tensor is a fiducial operator with spin $2$ and dimension $\widetilde{\Delta}_T = d- \Delta_T = 0$.

 The block associated with stress tensor exchanges will simply be written as $G_d^{(2)}(u,v)$. In terms of the OPE blocks, we can write the contribution from the stress tensor and its descendants to the four-point function as 
\begin{equation}
\label{eq:kinematicBlock}
   \langle V(x_1)V(x_2)W(x_3)W(x_4)\rangle \big{|}_{T} =   \frac{C_{VVT} C_{WWT}}{x_{12}^{2\Delta_V} x_{34}^{2\Delta_W}}   \; \big{\langle} \mathfrak{B}_T(x_1,x_2)\, \mathfrak{B}_T(x_3,x_4) \big{\rangle}_\text{phys.} 
\end{equation}
If we compute this two-point function of bilinears without doing the projection onto the physical part, we would again find a combination of the identity block and its shadow proportional to \eqref{eq:CPWdef}.

\vspace{10pt}
\subsection{A reformulation of the shadow operator formalism}

We will now explain our main results regarding a novel formulation of the shadow operator formalism for the global stress tensor block. The main technical observation will be the fact that the shadow of the stress tensor (which has vanishing conformal dimension) formally can be written as the descendant of a vector with dimension minus one. Subsequently we give an alternative interpretation of the vector as a ``soft'' diffeomorphism mode.

\paragraph{The stress tensor shadow as a total derivative.} Consider the shadow of the stress tensor:
\begin{equation}
\label{eq:QThigherd}
   \widetilde{T}^{\mu\nu}(x) \equiv  \frac{k_{d,2}}{\pi^{d/2}}
 \int d^d \xi \; I^{\mu\rho}(x-\xi) \; I^{\nu\sigma}(x-\xi)\;   T_{\rho\sigma} (\xi)\,,
\end{equation}
where $I^{\mu\rho}(x) \equiv \eta^{\mu\rho} - \frac{2}{x^2}\,x^\mu x^\rho$ is the inversion tensor. We observe that the operator $\widetilde{T}^{\mu\nu}$ can formally be written as a symmetric-traceless derivative:\footnote{  
See also \eqref{eq:Irelations} for useful relevant identities and p.\ 194 of \cite{Fradkin:1997df} for a more abstract way of writing this expression. The normalization in \eqref{eq:epsSdef} is chosen for later convenience.}
\begin{equation}
\label{eq:epsSdef}
\boxed{ \widetilde{T}_{\mu\nu}(x) \equiv 2C_T \, \frac{\pi^{d/2}}{k_{0,2}}\, 
   \mathbb{P}\,{}_{\mu\nu}^{\rho\sigma} \; \partial_\rho \epsS_\sigma(x)
\,,\qquad \epsS_\sigma(x) =
\frac{1}{2C_T}\frac{k_{0,2}k_{d,2}}{\pi^{d}} \int d^d\xi \; (x-\xi)^\alpha \, I^\beta_\sigma(x-\xi) \, T_{\alpha\beta}(\xi) }
\end{equation}
where we employ the symmetric-traceless projector 
\begin{equation}
 \mathbb{P}\,{}_{\mu\nu}^{\rho\sigma} = \frac{1}{2} \left( \delta_\mu^\rho \delta_\nu^\sigma + \delta_\nu^\rho \delta_\mu^\sigma \right) - \frac{1}{d} \, \eta_{\mu\nu} \, \eta^{\rho\sigma}\,.
\end{equation}
In other words, we formally write the stress tensor shadow as the descendent of a fiducial vector field with dimension $\Delta_\epsS = -1$ and spin $\ell_\epsS=1$. 
Our main task will be to study this mode in its own right. As we will argue, it allows for a useful reformulation of the technology we have described so far, and it relates it to an effective field theory of reparametrizations.

Eq.\ \eqref{eq:epsSdef} provides a nonlocal definition of $\epsS_\mu(x)$ in terms of the stress tensor. However, we would like to establish a local propagator for this mode. This can be achieved by observing that the two-point function  $\langle \widetilde{T}_{\mu\nu}\widetilde{T}_{\rho\sigma}\rangle$ can be written as a total derivative in much the same spirit as $\widetilde{T}_{\mu\nu}$ itself: by conformal invariance, we have\footnote{  We use a convention where 
$$
   \langle T_{\mu\nu}(x) T_{\rho\sigma}(y) \rangle  = 2 C_T \, {\mathbb P}_{\mu\nu}^{\alpha\gamma}  \;{\mathbb P}_{\rho\sigma}^{\beta\delta}  \; \left( \eta_{\alpha\beta} - 2 \, \frac{(x-y)_\alpha (x-y)_\beta}{(x-y)^2} \right) \left( \eta_{\gamma\delta} - 2 \, \frac{(x-y)_\gamma (x-y)_\delta}{(x-y)^2} \right) \frac{1}{(x-y)^{2d}} \,.
$$
}
\begin{equation}
\label{eq:TtTt}
\begin{split}
 &  \langle \widetilde{T}_{\mu\nu}(x) \widetilde{T}_{\rho\sigma}(y) \rangle  = 2 C_T \,\frac{k_{d,2}}{k_{0,2}}\,{\mathbb P}_{\mu\nu}^{\alpha\gamma}  \;{\mathbb P}_{\rho\sigma}^{\beta\delta}  \; \left( \eta_{\alpha\beta} - 2 \, \frac{(x-y)_\alpha (x-y)_\beta}{(x-y)^2} \right) \left( \eta_{\gamma\delta} - 2 \, \frac{(x-y)_\gamma (x-y)_\delta}{(x-y)^2} \right) \\
   &\qquad = -C_T\, \frac{k_{d,2}}{k_{0,2}}\, {\mathbb P}_{\mu\nu}^{\alpha\gamma}  \;{\mathbb P}_{\rho\sigma}^{\beta\delta}  \; \partial_\gamma^{(x)} \partial_\delta^{(y)} \left[ \left( \eta_{\alpha\beta} - 2 \, \frac{(x-y)_\alpha (x-y)_\beta}{(x-y)^2 } \right) \, (x-y)^2 \, \log \left( \mu^2 (x-y)^2 \right) \right] \,.
 \end{split}
\end{equation}
We interpret the expression in square brackets as the two-point function of $\epsS_\mu$:
\begin{equation}
\begin{split}
   & \langle \widetilde{T}_{\mu\nu}(x) \widetilde{T}_{\rho\sigma}(y) \rangle  = 4C_T^2 \, \frac{\pi^d}{k_{0,2}^2} \,  {\mathbb P}_{\mu\nu}^{\alpha\gamma}  \;{\mathbb P}_{\rho\sigma}^{\beta\delta}  \; \partial_\gamma^{(x)} \partial_\delta^{(y)} \langle \epsS_\alpha (x) \, \epsS_\beta(y) \rangle
\end{split}
\end{equation}
where we defined the ``logarithmic'' two-point function of the {\it stress tensor shadow mode} $\epsS_\mu$ as
\begin{equation}
\label{eq:epsPropNaive}
\boxed{
   \langle \epsS_\alpha(x) \epsS_\beta (y) \rangle 
   = - \frac{1}{C_T} \, \frac{k_{0,2}k_{d,2}}{4\pi^d}\, \left( \eta_{\alpha\beta} - 2 \, \frac{(x-y)_\alpha (x-y)_\beta}{(x-y)^2 } \right) \, (x-y)^2 \, \log \left[ \mu^2\, (x-y)^2  \right]
}
\end{equation}
Here, $\mu^2$ is an energy scale that we need to introduce to write a sensible expression. Note that this scale drops out after taking the symmetric-traceless derivatives -- formally it is an ambiguity in writing the above expression as a derivative. Similarly, any rescaling of the arbitrary scale $\mu^2$ results in the addition of analytic terms to the propagator, which drop out in physical quantities. We can view it as a hint that this theory is secretly a theory of the conformal anomaly. This connection will be made more precise in \S\ref{sec:W2Ward}.\footnote{  The necessity of introducing such a scale is also reminiscent of logarithmic conformal field theory \cite{Gurarie:1993xq,Gaberdiel:2001tr,Flohr:2001zs}: our field $\epsS_\mu$ looks much like a logarithmic primary. We thank D.\ Grumiller for pointing this out and note that it would be interesting to explore this further.}

Except for this arbitrary scale, one can write other ambiguity terms that leave \eqref{eq:TtTt} invariant, such as constant shifts of the propagator proportional to the inversion tensor $I_{\alpha\beta}$. These will turn out to be irrelevant for the physics we discuss. 

\paragraph{Coupling to external operators.} With the propagator \eqref{eq:epsPropNaive} at hand, we now need to understand the coupling to external probe operators. To this end, we observe that $\widetilde{T}_{\mu\nu}$ acts as a source for $T^{\mu\nu}$. We can thus use it to write the following dimensionless projector onto the conformal family of the stress tensor \cite{SimmonsDuffin:2012uy}:
\begin{equation}
\begin{split}
  |\widetilde{T}| & \equiv \frac{1}{2C_T} \frac{k_{0,2}}{\pi^{d/2}}
  \int d^d\xi \;  \widetilde{T}_{\mu\nu}(\xi)\, |0\rangle \langle0|\, {T}^{\mu\nu}(\xi) \\
  &= -
  \int d^d\xi \; \left\{ \frac{1}{d}\, \partial_\rho \epsS^\rho(\xi) \, |0\rangle \langle0|    T^{\mu}_\mu(\xi) + \epsS_\nu(\xi) \, |0\rangle \langle 0| \, \partial_\mu T^{\mu\nu}(\xi) \right\}
  \end{split}
  \end{equation}
where we used \eqref{eq:epsSdef} and integrated by parts the second term. 
We similarly define the projector $|T|$ such that we have the following properties: $(i)$ the projector is invariant under taking its shadow, $|\widetilde{T}| = |T|$;  $(ii)$ it squares to itself, $|T|^2 = |T|$; and $(iii)$ the insertion into correlation functions involving other stress tensors is trivial, $\langle T(x) \ldots \rangle = \langle T(x) |T| \cdots \rangle$. A few more details about these properties are given in Appendix \ref{sec:conventions}.

The stress tensor OPE block follows from applying this projector to a state associated with the insertion of a pair of external primaries $V$:
\begin{equation}
\label{eq:calc111}
\begin{split}
   |\widetilde{T}| V(x) V(y) \rangle &= - 
   \int d^d\xi \, \left\{ \frac{1}{d} \, |\partial_\rho \epsS^\rho(\xi) \rangle \langle T^\mu_\mu(\xi) V(x) V(y) \rangle + | \epsS_\nu(\xi) \rangle \; \partial_\mu \langle T^{\mu\nu}(\xi)V(x)V(y)  \rangle  \right\} \\
   &\equiv 
   \langle V(x) V(y) \rangle \times  |\mathcal{B}_{\epsS,V}^{(1)}(x,y)\rangle\,,
   \end{split}
\end{equation}
which defines the normalized bilinear $\mathcal{B}_{\epsS,V}^{(1)}(x,y)$ that describes the coupling of $\epsS^\mu$ to a bilinear pair of primaries.
We have factored out normalization factors in order to conform with conventions in lower dimensions.
The conformal Ward identity determines the divergence and the trace of the three-point function $ \langle T^{\mu\nu}(\xi)\,V(x)V(y)  \rangle$, which is nonzero due to contact terms. These can be parametrized as (see, for example, ref.\ \cite{Fradkin:1997df}
)\footnote{ 
There are well known ambiguities in the parametrization of these anomalies. Another common parametrization is \cite{Osborn:lectures}:
\begin{equation}
\begin{split}
      \langle T^\mu_\mu(\xi) V(x) V(y) \rangle &= -\Delta_V \,\left[ \delta^{(d)}(\xi-x) + \delta^{(d)}(\xi-y)\right] \, \langle V(x) V(y) \rangle \,,\\
     \partial_\mu \langle T^{\mu\nu}(\xi)\,V(x)V(y)  \rangle  &=- \left[ \delta^{(d)}(\xi-x)\,  \partial^\nu_{(x)}+ \delta^{(d)}(\xi-y) \,  \partial^\nu_{(y)}\right]  \; \langle V(x) V(y) \rangle \,.
 \end{split}
\end{equation}
Plugging this into \eqref{eq:calc111} gives the same result \eqref{eq:B1general}.
}
\begin{equation}
\label{eq:Ward1}
\begin{split}
      \langle T^\mu_\mu(\xi) V(x) V(y) \rangle &= 0  \,,\\
     \partial_\mu \langle T^{\mu\nu}(\xi)\,V(x)V(y)  \rangle  &= - \left[ \delta^{(d)}(\xi-x)\,  \partial^\nu_{(x)}+ \delta^{(d)}(\xi-y) \,  \partial^\nu_{(y)}\right]  \; \langle V(x) V(y) \rangle\\
    &\quad +\frac{\Delta_V}{d}  \partial^\nu_{(\xi)} \left[ \delta^{(d)}(\xi-x) + \delta^{(d)}(\xi-y)\right] \, \langle V(x) V(y) \rangle \,.
 \end{split}
\end{equation}
Plugging into \eqref{eq:calc111} gives
\begin{equation}
\begin{split}
 |\mathcal{B}_{\epsS,V}^{(1)}(x,y)\rangle
  &=  \frac{1}{\langle V(x) V(y) \rangle} \; \bigg\{ \frac{\Delta_V}{d} \, |\partial_\rho \epsS^\rho(x) + \partial_\rho \epsS^\rho(y) \rangle \langle V(x) V(y) \rangle \\
  &\qquad\qquad\quad\;\;\; +\left( | \epsS_\nu(x) \rangle \,  \partial^\nu_{(x)} +| \epsS_\nu(y) \rangle \, \partial^\nu_{(y)} \right) \langle V(x)V(y) \rangle \bigg\} 
   \end{split}
\end{equation}
Using the explicit form of the scalar two-point function, $\langle V(x) V(y) \rangle \propto (x-y)^{-2\Delta_V}$, we immediately obtain the bilinear coupling (which we now write as an operator identity):
\begin{equation}
\label{eq:B1general}
\boxed{
   {\cal B}^{(1)}_{\epsS,V}(x,y)    = \Delta_V \, \left\{ \frac{1}{d} \left( \partial_\mu \epsS^\mu(x) + \partial_\mu \epsS^\mu(y) \right) - 2\,  \frac{(\epsS(x)-\epsS(y))^\mu  \,(x-y)_\mu}{ (x-y)^2 } \right\}
}
\end{equation}
Note that the only way $\mathcal{B}_{\epsS,V}^{(1)}(x,y)$ depends on the operator $V$ is through an overall factor $\Delta_V$. We decide to keep this factor as part of the definition of the bilocal.

Our proposal is that this bilinear of local fields $\epsS^\mu(x)$ is identical to the stress tensor OPE block:\footnote{ Note that the OPE coefficient $C_{VVT}$ is fixed by the conformal Ward identity: $C_{VVT}= \frac{\Delta_V \, d \, \Gamma\left(\frac{d}{2}\right)}{\pi^{d/2} (d-1)}$.}
\begin{equation}
\label{eq:Bequal}
\boxed{ {\cal B}^{(1)}_{\epsS,V}(x,y) =C_{VVT}  \; \mathfrak{B}_T(x,y).
}
\end{equation}
Both objects are bilinear operators. The l.h.s.\ contains local insertions of $\epsS^\mu$; we think of it as the vertex coupling bilinear operators to the mode $\epsS_\mu$.  The r.h.s.\ is the OPE block (i.e., a nonlocal smearing of the stress tensor). In order to make sense of this equation, one needs to know the propagator of $\epsS^\mu$, which we provided in \eqref{eq:epsPropNaive}. Inside correlation functions, one can then check the identity.

\paragraph{Application to four-point functions.}
We can also use the shadow operator formalism directly in correlation functions to derive the bilinear couplings of the mode $\epsS^\mu$. To this end, we insert both of the projectors $|T|$ and $|\widetilde{T}|$ in a four-point function:
\begin{equation}
\begin{split}
   &\langle V(x_1)V(x_2)W(x_3)W(x_4) \rangle = \langle V(x_1)V(x_2)|T|\widetilde{T}|W(x_3)W(x_4) \rangle_{\text{phys.}} \\
   &\qquad = \frac{1}{4C_T^2} \frac{k_{0,2}^2}{\pi^d}
    \iint d^d\xi \,d^d\xi' \; \langle V(x_1)V(x_2) T^{\mu\nu}(\xi) \rangle \langle \widetilde{T}_{\mu\nu}(\xi) \widetilde{T}_{\rho\sigma}(\xi') \rangle \langle T^{\rho\sigma}(\xi') W(x_3)W(x_4) \rangle_{\text{phys.}}
\end{split}
\end{equation}
Writing both instances of $\widetilde{T}$ as descendants of $\epsS$, we can then follow a similar procedure as in the previous paragraph: we integrate both derivatives by parts and use the conformal Ward identity \eqref{eq:Ward1} twice to localize the integrals. As a result, the above expression reduces to a two-point function of bilinears:
\begin{equation}
   \langle V(x_1)V(x_2)W(x_3)W(x_4) \rangle \big{|}_T =     \langle VV \rangle \langle WW\rangle \times \langle {\cal B}^{(1)}_{\epsS,V}(x_1,x_2){\cal B}^{(1)}_{\epsS,W}(x_3,x_4) \rangle_{\text{phys.}}
\end{equation}
This shows very directly that the global stress tensor block should be computed by the two-point function of ${\cal B}_{\epsS,V}^{(1)}$, which is just a simple linear combination of $\epsS$-propagators. By comparing with \eqref{eq:kinematicBlock} we also verify \eqref{eq:Bequal}. We will now investigate these claims in more detail. 

\vspace{10pt}
\subsection{Conformal partial waves in arbitrary even dimensions}

As the most immediate application of our reformulation of the shadow operator formalism, we will now illustrate how to compute global conformal blocks. We already know from \eqref{eq:GfromKin} that these should be computed by two-point functions of OPE blocks. Note that we will work in Euclidean signature, so we expect to find conformal partial waves.

Let us compute the two-point function of bilinears, using the ``vertices'' \eqref{eq:B1general} and the propagator \eqref{eq:epsPropNaive}. No conformal integrals need to be done. The calculation simply consists of taking a linear combination of the $\epsS^\mu$-propagator and its derivatives. This straightforward calculation yields:
\begin{equation}
\label{eq:globalBlock}
\boxed{
  \big{\langle} {\cal B}^{(1)}_{\epsS,V}(x_1,x_2) \, {\cal B}^{(1)}_{\epsS,W}(x_3,x_4) \big{\rangle} = -\frac{8\Delta_V \Delta_W}{C_T} \,  \frac{k_{0,2}k_{d,2}}{4\pi^d} \, \left(  \frac{4}{d} +  \frac{1-v}{u} \; \log v \right)
  }
\end{equation}
which is written in terms of the cross ratios \eqref{eq:CRs}. {\it We claim that \eqref{eq:globalBlock} is proportional to the global stress tensor conformal partial wave $f_{\Delta=d}^{(\ell=2)}$ in arbitrary even dimension}. In odd dimensions \eqref{eq:globalBlock} corresponds to just the shadow block. We will later see in more detail why a soft mode theory of physical conformal blocks is difficult to obtain in odd dimensions. Note that \eqref{eq:globalBlock} is simpler than perhaps expected. Let us therefore understand it in more detail. 

As a first consistency check, note that \eqref{eq:globalBlock} is an eigenfunction of the Casimir \eqref{eq:Casimir} with eigenvalue $-2d$, as expected for any linear combination of stress tensor block and shadow block. Note further that \eqref{eq:globalBlock} is not just any such linear combination but a very particular one which has the feature that it is single-valued (since $0 < u,v < 1$).

\paragraph{Comparison with known blocks in even dimensions.}

Let us now compare with known results for global conformal partial waves in various dimensions, distinguishing even and odd dimensions. For even dimensions, these were computed in \cite{Dolan:2003hv,Dolan:2011dv} for more general situations (arbitrary internal and external operators) in terms of hypergeometric functions. Note that these objects are increasingly complicated in higher dimensions. Closed form expressions are available in $d=2,4,6$ in \cite{Dolan:2003hv}. In the special case of the stress tensor block for pairwise equal external operators, their results are indeed proportional to \eqref{eq:globalBlock}. For instance, in $d=2,4$ the blocks and shadow blocks associated with stress tensor exchanges between pairwise equal primary operators take the following form (the case of $d=6$ is similar):
\begin{equation} 
\label{eq:4dblocks}
\begin{split}
\triangleright \;\,d=2:\quad  \text{ physical block:}\quad  G_2^{(2)} & = 3\,\left[ \frac{z-2}{z} \, \log (1-z) - 2 \right ]+ \text{c.c.} \\
  \text{ shadow block:} \quad  G_0^{(2)} &=  \frac{1}{4}\,\left[ \frac{z-2}{z}  \, \log (1-\bar{z})  \right ]+ \text{c.c.}\\
\triangleright \;\,d=4:\quad  \text{ physical block:}\quad G_4^{(2)} &=   10 \,\left[ - \frac{ \bar{z}}{z}\; \frac{z^2 - 6z + 6}{(z-\bar{z})}  \, \log (1-z) +3\right]+ \text{c.c.} \\
  \text{ shadow block:} \quad  G_0^{(2)} &= \frac{1}{18} \, \left[\frac{ \bar{z}}{z}\; \frac{z^2 - 6z + 6}{(z-\bar{z})}  \, \log (1-\bar{z})  \right ]+ \text{c.c.}
\end{split}
\end{equation}
where  $u \equiv z \bar{z}$ and $v \equiv (1-z)(1-\bar{z})$ and ``$c.c.$'' denotes complex conjugation (recall that in Euclidean signature, $\bar{z} \equiv z^*$). One can readily verify the short distance behavior \eqref{eq:shortdist} and the fact that the CPW \eqref{eq:CPWdef} is proportional to \eqref{eq:globalBlock}.

Compared to the very complicated and strongly dimension-dependent general results of \cite{Dolan:2011dv}, our expression \eqref{eq:globalBlock} for the stress tensor CPW is remarkably simple. This simplicity of the stress tensor partial wave -- and the existence of a simple expression that is valid in any dimension -- is a new result as far as we are aware.

\vspace{10pt}
\subsection{Generalization: conserved currents with higher spin}
\label{sec:higherspin}

Our discussion focuses on the OPE block associated with the stress tensor. While it is unclear how to generalize it to OPE blocks of arbitrary internal primaries, the generalization to conserved currents is straightforward. Consider a symmetric-traceless conserved current $J_{\mu_1 \cdots \mu_\ell}$ (i.e., $\Delta_J = d+\ell-2$). Its shadow is 
\begin{equation}
\label{eq:Jshadow}
   \widetilde{J}^{\mu_1 \cdots \mu_\ell}(x) \equiv \frac{k_{\Delta_J,\ell} }{\pi^{d/2}} \int d^d \xi \; \frac{1}{(x-\xi)^{2(2-\ell)}} \left( \prod_{i=1}^\ell \;  I^{\mu_i \nu_i}(x-\xi) \right) J_{\nu_1\cdots \nu_\ell} (\xi) \,,
\end{equation}
where the product of inversion tensors is understood to inherit symmetries and tracelessness from contraction with the current.
As one easily verifies, this can be written as a symmetric-traceless derivative:
\begin{equation}
\label{eq:Jderiv}
\begin{split}
  \widetilde{J}_{\mu_1 \cdots \mu_\ell} &= \frac{k_{\Delta_J,\ell} }{\pi^{d/2}} \; \left(\partial^{\,}_{(\mu_1} \epsS^{(J)}_{\mu_2 \cdots \mu_\ell)} - \text{traces} \right) \,,\\
\text{where}\quad  \epsS^{(J)}_{\mu_2 \cdots \mu_\ell} (x)&= \int d^d \xi \; \frac{(x-\xi)^{\nu_1}}{(x-\xi)^{2(2-\ell)}} \left( \prod_{i=2}^\ell \;  I_{\mu_i}^{\nu_i}(x-\xi) \right) J_{\nu_1\cdots \nu_\ell} (\xi) \,.
  \end{split}
\end{equation}
where the bracket around indices means complete symmetrization, and we subtract traces such as to obtain an object that is traceless in any two indices. 

Starting from the observation that the shadow of any symmetric-traceless conserved current can be written as a descendant of some fiducial field $\epsS^{(J)}$ with $(\Delta_{\epsS^{(J)}},\ell_{\epsS^{(J)}}) = (1-\ell_J,\ell_J-1)$, one could formulate a theory similar to the one for our reparametrization mode. Note that this can be thought of as taking a result of \cite{deBoer:2016pqk} to the next level: there it was observed that the smeared representation of OPE blocks for conserved currents localizes onto just a time slice of the full causal diamond. Our result shows that the OPE blocks for conserved currents allow for an even more local description: they can be computed in terms of reparametrization modes (or other symmetry generating modes), which completely localize to the two insertion points of the original operators in the OPE. In other words, the computation of global conformal blocks in these cases reduces to evaluating a local exchange diagram of symmetry generating modes $\epsS^{(J)}_{\mu_2 \cdots \mu_\ell}$. This makes the evaluation of any conformal integrals unnecessary and should simplify various calculations in the literature. 

We will not pursue this more general case further. For clarity of presentation, we have only discussed the stress tensor case with $\ell=2$. However, from \eqref{eq:Jderiv} it is clear that it should be possible to generalize our entire discussion to conserved currents with higher spin. It would be interesting to see if a simple expression for the associated conformal partial waves in arbitrary dimension can be obtained, analogous to \eqref{eq:globalBlock}.

\vspace{10pt}
\section{A theory of reparametrization modes in higher dimensions}
\label{sec:reparametrizations}

We have identified a soft mode $\epsS_\mu$ in higher dimensional CFT, and its coupling to matter. These ingredients are related to the shadow of the energy momentum operator in the manner described, and were sufficient to elucidate the connection between propagation of the soft mode between bilinear operators and the computation of global conformal blocks or conformal partial waves.

We will now explore further the observation that the soft mode can equivalently be understood as a reparametrization mode that sources the stress tensor. Using this insight, in this section we take preliminary steps towards writing an effective action for this mode which reproduces the above results.  Specifically, we will discuss the linearized action, which is sufficient to reproduce the above results. The full non-linear action will be the subject of a separate, future discussion. 

\vspace{10pt}
\subsection{Effective action from the conformal Ward identity} 
\label{sec:W2Ward}

Consider an infinitesimal reparametrization $x^\mu \rightarrow x^\mu +  \epsilon^\mu (x)$ which provides a source for the stress tensor via $\delta g_{\mu\nu} = -2 \partial_{(\mu} \epsilon_{\nu)}$. In the special case where $\epsilon^\mu(x)$ is a conformal Killing vector, we can subsequently perform a conformal transformation to remove this source. In order to make this manifest, we work with the source $\delta' \! g_{\mu\nu} = -2 \partial_{(\mu} \epsilon_{\nu)} + \frac{2}{d}  \, \eta_{\mu\nu} (\partial. \epsilon)$, which vanishes on conformal transformations. Of course, we recognize this combination as being of exactly the same form as the shadow of the stress tensor, see \eqref{eq:epsSdef}. But now, we wish to study the dynamics of the {\it reparametrization mode} $\epsilon^\mu(x)$ by thinking of it as a source for the stress tensor in the spirit of effective field theory. To this end, we work with the connected generating functional $W[\epsilon]$, which is defined in terms of the partition function of the Euclidean CFT, $Z_0 = \int [d\Phi] \; e^{-S_{CFT}}$, as
\begin{equation}
  e^{-W[\epsilon] }=  \frac{1}{Z_0}\int [d\Phi] \; e^{-S_{CFT} - \int \frac{1}{2} \,\delta'\! g_{\mu\nu} T^{\mu\nu}}   \,.
\end{equation}
Let us use a convention where $T^{\mu\nu}$ is traceless and symmetric.
The quadratic action for the reparametrization mode then reads as
\begin{equation}
    W_2[\epsilon] = -\frac{1}{2}\int d^dx \, d^dy \; \partial_{\mu} \epsilon_{\nu}(x) \; \partial_{\rho} \epsilon_{\sigma}(y) \; \langle T^{\mu\nu}(x) T^{\rho\sigma}(y) \rangle_\text{conn.} \,.
\end{equation}
We wish to integrate by parts one of the derivatives and use the conformal Ward identity in even dimensions \cite{Fradkin:1996is}:
\begin{equation}
\label{eq:dTT}
\begin{split}
   \partial_\mu \langle T^{\mu\nu}(x) T^{\rho\sigma}(y) \rangle_\text{conn.} = C_T\,n_d \,\bigg\{ \partial^\nu \partial^\rho \partial^\sigma - \frac{d-1}{d} \, \eta^{\nu(\rho} \partial^{\sigma)} \, \Box - \frac{1}{d^2} \, \eta^{\rho\sigma} \partial^\nu\Box \bigg\} \, \Box^{\frac{d-2}{2}} \delta^{(d)}(x-y) \,,
\end{split}
\end{equation}
where $n_d = - 2^{2-d}\, \pi^{d/2}/ \Gamma(d+2) \Gamma(\frac{d}{2})$.
Note that in odd dimensions there is no such anomalous contribution and the quadratic action we define here simply vanishes. This should be tied to the fact that the conformal blocks are more complicated in odd dimensions.
Plugging \eqref{eq:dTT} into the action $W_2$ gives:
\begin{equation}
\label{eq:W2full}
\boxed{
   W_2[\epsilon] = -C_T\,n_d \, \frac{d-1}{4d}  \int d^dx \;\, \epsilon^\mu(x)  \left( \eta_{\mu\nu} \, \Box - \frac{d+2}{d}  \, \partial_\mu\partial_\nu \right) \, \Box^{\frac{d}{2}}\ \epsilon^\nu(x) 
   }
\end{equation}
We can now discuss the resulting propagator for our soft mode. In order to invert the kernel, we can pass to momentum space where $\partial_\mu \rightarrow i k_\mu$. The momentum space propagator follows immediately after inverting the integration kernel: 
\begin{equation}
\label{eq:epsepsMom}
  \langle \epsilon^\mu(k) \epsilon^\nu(-k) \rangle = - \frac{1}{C_T\,n_d} \, \frac{2d}{d-1} \, \left( \eta^{\mu\nu} \, k^2 - \frac{d+2}{2} \, k^\mu k^\nu \right) (-k^2)^{-\frac{d+4}{2}}\,.
\end{equation}
Note that the scaling $ (-k^2)^{-\frac{d+2}{2} }$ is what one would formally expect for the momentum space two-point function of dimension-$(-1)$ operators \cite{Bzowski:2013sza}. 
We will make this more precise now.
In fact, by Fourier transforming back to position space, we can recover the result of our reformulation of the shadow operator formalism! Let us now see how this works. 

The Fourier transform of \eqref{eq:epsepsMom} is naively divergent, so we need to regularize it. This is done by first reducing the divergence for small momenta:
\begin{equation}
\begin{split}
  \langle \epsilon^\mu(x) \epsilon^\nu(0) \rangle 
  &=- \frac{1}{C_T\,n_d} \, \frac{2d}{d-1} \int \frac{d^dk}{(2\pi)^d} \; e^{ikx} \;  \left( \eta^{\mu\nu} \, k^2 - \frac{d+2}{2} \, k^\mu k^\nu \right) (-k^2)^{-\frac{d+4}{2}} \\
    &=- \frac{1}{C_T\,n_d} \, \frac{1}{2(d-1)}\int \frac{d^dk}{(2\pi)^d}  \; e^{ikx} \;  \left( \eta^{\mu\nu} \, \Box_{(k)} - 2 \, \partial^\mu_{(k)} \partial^\nu_{(k)} \right) (-k^2)^{-\frac{d}{2}} \\
        &=\frac{1}{C_T\,n_d} \, \frac{1}{2(d-1)}  \left( \eta^{\mu\nu} \, x^2- 2 \, x^\mu x^\nu \right)\int \frac{d^dk}{(2\pi)^d}  \; e^{ikx} \;   (-k^2)^{-\frac{d}{2}} \,,
\end{split}
\end{equation}
where we integrated by parts the $k$-derivatives thus converting them into a position space tensor structure. In the last line we recognize the inversion tensor $I_{\mu\nu}(x) = \eta_{\mu\nu} - 2\, \frac{x^\mu x^\nu}{x^2}$.

Next, we use differential regularization \cite{Freedman:1991tk,Latorre:1993xh} to replace $|k|^{-d}$ by a regulated expression with a well-defined Fourier transform:
\begin{equation}
  |k|^{-d} \Big{|}_\text{reg.} \equiv -\frac{1}{2(d-2)} \, \Box_{(k)} \left( \frac{\log(|k|^2 \mu^{-2})}{|k|^{d-2}} \right) \,,
\end{equation}
where $\Box_{(k)}$ can again be replaced by a factor of $(-x^2)$, using integration by parts. In writing the regularized expression it is necessary to introduce an arbitrary energy scale $\mu^2$. We note the following Fourier transform of the regulated expression: 
\begin{equation}
\int \frac{d^dk}{(2\pi)^d} \; e^{ikx} \, \frac{\log(|k|^2 \mu^{-2})}{|k|^{d-2}} = -\frac{d-2}{2^{d-1} \pi^{d/2} \Gamma(\frac{d}{2})}\, \frac{1}{x^2} \left[ \log \left( \mu^2x^2 \right) - \log 4 + \gamma  - \psi\left(\frac{d-2}{2} \right) \right]
\end{equation}
We can clearly drop the three constant terms in the square bracket as these can be absorbed into a redefinition of the arbitrary scale $\mu$ (in abuse of notation, we will denote the redefined $\mu$ by the same letter).
Putting all these pieces together, we find: 
\begin{equation}
\label{eq:epsepsR}
\begin{split}
  \langle \epsilon^\mu(x) \epsilon^\nu(0) \rangle 
               = -\frac{1}{C_T} \,  \frac{k_{0,2} k_{d,2}}{4\pi^d} \,  I^{\mu\nu}(x)\, x^2 \;\log \left( \mu^2x^2 \right) 
\end{split}
\end{equation}
This is precisely the propagator \eqref{eq:epsPropNaive} that we derived from the shadow operator formalism upon identifying the reparametrization mode $\epsilon^\mu$ in terms of the mode $\epsS^\mu$ whose descendant is the stress tensor shadow! At least at the level of two-point functions, we conclude that the reparametrization mode and shadow mode are the same:
\begin{equation}
\epsilon^\mu(x) = \epsS^\mu(x) \,.
\end{equation}
In the following we argue for this relation more generally.

\vspace{10pt}
\subsection{Monodromy projection and symmetries of the quadratic action}
We have now established that the reparametrization modes $\epsilon^\mu$ exhibit the same dynamics as the shadow modes $\epsS^\mu$ whose derivative is the shadow of the stress tensor. However, we are not done yet, as we still need to implement the monodromy projection onto the physical contributions to correlation functions. We now argue that the monodromy projection can be understood in terms of the symmetries of the effective action.

Note that the quadratic action \eqref{eq:W2full} can be written as follows:
\begin{equation}
\label{eq:W2new}
   W_2[\epsilon] = -\frac{C_T\,n_d}{2} \, \int d^dx \; \mathbb{P}_{\mu\nu}^{\rho\sigma} \partial_\rho \epsilon_\sigma(x) \; \,\left[ \partial^\lambda\partial^{(\mu} - \frac{d-1}{d} \,  \Box\,\eta^{\lambda(\mu}  \right]  \partial^{\nu)} \,\Box^{\frac{d-2}{2}} \epsilon_\lambda(x)\,.
\end{equation}
This way of writing the action makes the connection with the shadow operator, as well as some important symmetry properties manifest. The first factor ($\mathbb{P}_{\mu\nu}^{\rho\sigma} \partial_\rho \epsilon_\sigma$) is just the symmetric traceless source associated with infinitesimal reparametrizations, $-\frac{1}{2} \, \delta'\! g_{\mu\nu}$. In other words, it is proportional to the longitudinal shadow operator dual to the stress tensor, c.f., \eqref{eq:epsSdef}. The second factor in \eqref{eq:W2new} can be thought of as the stress tensor itself, expressed as an operator in our effective field theory. Indeed, we have from the definition of the shadow transform:
\begin{equation}
\label{eq:Teps}
\begin{split}
   T^{\rho\sigma}(x) &\equiv \doublewidetilde{{T}}{}^{\rho\sigma}(x) \equiv \frac{k_{0,2}}{\pi^{d/2}} \, \frac{1}{2C_T} \int d^d y \; \langle T^{\rho\sigma}(x) T^{\mu\nu}(y) \rangle \, \widetilde{T}_{\mu\nu}(y) \\
   &= C_T\,n_d \; \mathbb{P}^{\rho\sigma}_{\mu\nu} \left[ \partial^\lambda\partial^{(\mu} - \frac{d-1}{d} \,  \Box\,\eta^{\lambda(\mu}  \right]  \partial^{\nu)} \,\Box^{\frac{d-2}{2}} \epsS_\lambda(x) \,,
\end{split}
\end{equation}
where we wrote $\widetilde{T}_{\mu\nu}$ in terms of $\epsS^\mu$ using \eqref{eq:epsSdef}, and then integrated by parts using the Ward identity \eqref{eq:dTT}. In short, the action \eqref{eq:W2new} can be simply understood as the coupling $W_2[\epsilon] \propto \int \delta'\! g_{\mu\nu} \; T^{\mu\nu}[\epsilon]$, where $T^{\mu\nu}[\epsilon]$ should be understood as in \eqref{eq:Teps}.
 This is, of course, not surprising, but it is useful to think about this standard coupling in terms of the reparametrization modes as it makes the connection with shadow operators very clear.

\paragraph{Symmetries of the quadratic action.} The action \eqref{eq:W2new} is useful for analyzing the symmetries. The two factors corresponding to the shadow of the stress tensor and the stress tensor itself, give rise to two sets of symmetries $\epsilon_\mu \rightarrow \epsilon_\mu + \delta \epsilon_\mu$:
\begin{itemize}
   \item The source/shadow of the stress tensor $(-\frac{1}{2} \, \delta'\! g_{\mu\nu})$ is invariant under:
   \begin{equation}
      \delta \epsilon_\mu = K_\mu \quad \text{with} \quad \partial_{(\mu} K_{\nu)} - \frac{1}{d} \, \eta_{\mu\nu} (\partial.K) = 0 \,,
   \end{equation}
   that is, $\delta\epsilon_\mu$ is a conformal Killing vector. There are the usual $\frac{1}{2}(d+1)(d+2)$ solutions to this, which are a manifestation of the $SO(d+1,1)$ conformal symmetry:
   \begin{equation}
      K_\mu(x) = a_\mu - \omega_{[\mu\nu]} x^\nu + \lambda \, x^\mu +  x^2 I_{\mu\nu}(x) \, b^\nu \,, \;\;\qquad I_{\mu\nu} = \eta_{\mu\nu} - 2 \, \frac{x_\mu x_\nu}{x^2} \,,
   \end{equation}
   which parametrize translations ($a^\mu$), rotations ($\omega_{[\mu\nu]}$), scale transformations ($\lambda$), and special conformal transformations ($b^\mu$).
   \item The stress tensor part $T^{\mu\nu}$ as in \eqref{eq:Teps} is invariant under $\delta\epsilon_\mu$ with
   \begin{equation}
   \label{eq:symmPhys}
      \mathbb{P}^{\rho\sigma}_{\mu\nu} \left[ \partial^\lambda\partial^{(\mu} - \frac{d-1}{d} \,  \Box\,\eta^{\lambda(\mu}  \right]  \partial^{\nu)} \,\Box^{\frac{d-2}{2}} \, \delta\epsilon_\lambda = 0 \,.
   \end{equation}
   We refer to these symmetries as {\it conformal redundancies}. They represent an invariance of the definition of the energy-momentum tensor in terms of the reparametrization mode. 
\end{itemize}

\paragraph{Projecting out the shadow modes.} 

We have now reached the point where we can present a proposal for separating the physical block from the unphysical shadow block, at the level of our effective field theory. In order for the reparametrization mode to describe exchanges of the physical stress tensor, but not of its shadow operator, we need to supplement our effective field theory with an appropriate gauge symmetry, such that the physical stress tensor is a gauge invariant operator but its shadow is not.\footnote{  We thank Kristan Jensen for useful conversations on this issue.} In other words,  we now wish to gauge precisely the {\it conformal redundancies} \eqref{eq:symmPhys}. Correspondingly, the propagation of the shadow operator in physical correlation functions will be forbidden. {\it In low-dimensional cases we verify below that this corresponds to the prescription of omitting gauge modes in the quadratic action.}

Note that the stress tensor can be written as a convolution of its shadow, with a known kernel which is proportional to the 2-point function $\langle T_{\mu \nu} T_{\rho \sigma}\rangle$, see \eqref{eq:Teps}. The conformal redundancies can be characterized equivalently as zero modes of that kernel -- those are symmetries which act non-trivially on the shadow yet leave the result of the convolution, the stress tensor, invariant. While we leave a complete characterization of the conformal redundancies (especially for the non-linear action), for future work, this suggest a direct connection with the phenomenon of pole skipping (to be discussed further below).

\vspace{10pt}
\subsection{Coupling reparametrizations to external operators}
Obviously, it is also possible to derive the bilinear couplings to external operators, \eqref{eq:B1general}, from the reparametrization mode technique. This strategy is the immediate generalization of the way matter fields are coupled to the Schwarzian mode in one dimension \cite{Almheiri:2014cka,Maldacena:2016hyu,Maldacena:2016upp}, and holomorphic reparametrizations are coupled to primary operators in CFT$_2$ \cite{Haehl:2018izb,Cotler:2018zff}: we start with the conformal two-point function and perform a reparametrization. To linear order in the reparametrization, we obtain the bilinear coupling \eqref{eq:B1general}.

Let us make this more precise.
We start with the coupling of the probe operators to sources, which is invariant under coordinate transformations $x^\mu \rightarrow x'{}^\mu$. It follows that
\begin{equation}
\begin{split}
   \int d^dx \,d^dy \, \langle V(x) V(y) \rangle J(x) J(y )
   &=  \int d^dx' d^dy' \; \langle V(x') V(y') \rangle \; J(x') J(y' ) \\
   &=  \int d^dx \,d^dy \; \langle V(x') V(y') \rangle \left[ \Omega(x) \Omega(y)\right]^{\frac{\Delta}{2}}  J(x) J(y ) 
   \end{split}
\end{equation} 
where $\Omega(x)$ is the conformal factor, related to the Jacobian by $ \big| \frac{\partial x'}{\partial x} \big| = \Omega^{d/2}$, and we also used $J(x') = J(x) \, \Omega^{(\Delta-d)/2}$.
From this we can read off the full nonlinear coupling of finite reparametrizations to pairs of scalar operators $V$. It is given by the reparametrized two-point function:
\begin{equation}
\label{eq:Bnonlinear}
   \overline{\cal B}_V(x,y) \propto    \left| \frac{\partial x'}{\partial x} \right|^{\frac{\Delta}{d}} \; \left| \frac{\partial y'}{\partial y} \right|^{\frac{\Delta}{d}} \, \frac{1}{\left( x'(x) - y'(y) \right)^{2\Delta} }  
\end{equation}
These objects are natural bilinear operators in the theory (local operators in kinematic space).

An infinitesimal reparametrization $x^\mu \rightarrow x'^\mu = x^\mu + \epsilon^\mu(x)$ gives a 
conformal factor  $\Omega(x) = 1 + \frac{2}{d} \, \partial_\rho \epsilon^\rho(x)$. This is related to the Jacobian of the transformation by $\big| \frac{\partial x'}{\partial x} \big| = 1 + \partial_\rho \epsilon^\rho$.
The perturbative couplings of infinitesimal reparametrizations follow from this structure by expanding \eqref{eq:Bnonlinear} in $\epsilon$: 
\begin{equation}
\overline{\cal B}_{\epsilon,V}(x,y)  =  \langle V(x) V(y) \rangle \, \left[ 1 +  {\cal B}^{(1)}_{\epsilon,V}(x,y) + {\cal B}^{(2)}_{\epsilon,V}(x,y) + \ldots \right]
\end{equation}
where ${\cal B}^{(1)}_{\epsilon,V}(x,y)$ is the same object as derived in \eqref{eq:B1general} where we identify the logarithmic primary $\epsS_\mu$ with the reparametrization mode $\epsilon_\mu$.
By expanding \eqref{eq:Bnonlinear} to higher orders, we obtain higher order couplings involving $n$ instances of $\epsilon_\mu$. The coupling at $n$-th order in $\epsilon_\mu$ is of the form 
\begin{equation}
\label{eq:Bp}
  {\cal B}_{\epsilon,V}^{(n)} = \frac{1}{n!} \left(  {\cal B}_{\epsilon,V}^{(1)} \right)^n + \text{lower order in } \Delta_V \,.
\end{equation}
 For $n=1$ these bilinears describe the physics of a single stress tensor exchange in the OPE. This is the leading answer at large $C_T$.  What is the meaning of higher order couplings ${\cal B}_{\epsilon,V}^{(n)}$? In two-dimensional CFTs, it is natural to expect that multi-$\epsilon$ exchanges correspond to Virasoro contributions to the conformal block. An example of a calculation where \eqref{eq:Bp} was used for all $n$ (and ``lower orders in $\Delta_V$'' are not needed), is the exponentiation of the light-light block, where all external operators are light compared to the central charge \cite{Fitzpatrick:2014vua}. This was verified for $d=2$ in \cite{Cotler:2018zff} using reparametrization mode techniques. In that case ladder exchange diagrams of reparametrization modes dominate. We review the two-dimensional calculation in Appendix \ref{sec:LLLL}. It would be interesting to explore similar statements in higher dimensions, where one might expect a similar exponentiation to occur \cite{Maxfield:2017rkn}. 
In higher dimensions, we expect that multi-$\epsilon$ diagrams compute multi-stress tensor exchanges. These are not universal in $d>2$, but there might well exist certain contributions (such as ``lowest twist'' contributions) or kinematic regimes where a universal answer exists. Such exchanges were recently considered in \cite{Fitzpatrick:2019zqz,Fitzpatrick:2019efk,Kulaxizi:2019tkd,Huang:2019fog}. It would be interesting to explore the relation with our formalism.

\vspace{10pt}
\section{Two-dimensional CFTs}
\label{sec:2d}

In this section we investigate in more detail the two-dimensional case, expanding on previous results of \cite{Haehl:2018izb,Cotler:2018zff} (see also \cite{Turiaci:2016cvo}).
In $d=2$ we use complex coordinates $(z,\bar{z})$. Most of our discussion will be in Euclidean signature, where $\bar{z} = z^*$. For more details about conventions, see Appendix \ref{sec:conventions}.

\vspace{10pt}
\subsection{Zero temperature}

In $d=2$, the stress tensor has components $T \equiv -2\pi T_{zz}$ and $\overline{T} \equiv -2\pi T_{\bar z \bar z}$. Similarly, the shadow of the stress tensor has non-zero components $\widetilde{T} \equiv -2\pi \widetilde{T}^{ z  z}$ and $\widetilde{\overline{T}} \equiv -2\pi \widetilde{T}^{\bar z\bar z}$. 
The shadow of the holomorphic stress tensor can be written as follows:
\begin{equation}
    \widetilde{T} (z,\bar{z})  =  \frac{2}{\pi} \int d^2z'  \; \frac{(z-z')^2}{(\bar{z}-\bar{z}')^2} \; {T}({z}') 
\end{equation}
and similarly for the shadow $\widetilde{\overline{T}}$ of the anti-holomorphic stress tensor $\overline{T}$. It is straightforward to write these shadow currents as symmetric-traceless derivatives:
\begin{equation}
\label{eq:2dshadow}
   \widetilde{{T}} = -\frac{c}{3} \, \bar{\partial} \epsS \qquad \text{ with }\qquad \epsS \equiv  \frac{6}{\pi\,c} \int d^2z'  \; \frac{(z-z')^2}{(\bar{z}-\bar{z}')} \; T(z') \,,
\end{equation}
and similarly for $\widetilde{\overline{T}} = -\frac{c}{3} \partial \overline{\epsS}$. The central charge appearing in \eqref{eq:2dshadow} is related to the usual central charge in two dimensions by $c=4\pi^2 C_T$.
Instead of shadow operators, we can think, as discussed above, in terms of holomorphic and anti-holomorphic reparametrizations $(z,\bar{z}) \rightarrow (z+\epsilon,\bar{z}+\bar{\epsilon})$. 

The reparametrization modes have an action as described in \S\ref{sec:reparametrizations}. The $d=2$ version of the Euclidean propagator \eqref{eq:epsepsR} (or \eqref{eq:epsPropNaive}) takes the following form:
\begin{equation}
\label{eq:2dprop}
  {\cal G}_\epsS^E (1,2) \equiv  \langle \epsS(z_1,\bar{z}_1) \epsS(z_2,\bar{z}_2) \rangle = \frac{6}{c} \, (z_1-z_2)^2 \;  \log \big[ \mu^2 (z_1-z_2) (\bar{z}_1 - \bar{z}_2) \big] 
\end{equation}
There is a similar expression for $\langle\overline{\epsS}\overline{\epsS}\rangle$, while mixed propagators vanish, $\langle \epsS \, \overline{\epsS} \rangle = 0$.
In the same way, we can use the bilinear coupling \eqref{eq:B1general}, viz.,
\begin{equation}
\label{eq:2dbilinear}
   {\cal B}_{\epsS,V}^{(1)}(z_1,\bar{z}_1;z_2,\bar{z}_2) =  \frac{\Delta_V}{2}\left[  \partial \epsS_1 + \partial \epsS_2 -2\, \frac{\epsS_1-\epsS_2}{z_1-z_2} \right]  + \text{anti-holo.}\,,
\end{equation}
to compute the conformal partial wave in $d=2$. We can further exploit holomorphic factorization in two dimensions by recognizing ${\cal B}_{\epsS,V}^{(1)}$ as the sum of two decoupled holomorphic and anti-holomorphic terms. This is useful for coupling the fields $\epsS$ and $\overline{\epsS}$ to spinning primaries with holomorphic and anti-holomorphic dimensions $(h,\bar{h})$:
\begin{equation}
{\cal B}_{\epsS,h}^{(1)}(1,2) \equiv h \, \left[  \partial \epsS_1 + \partial \epsS_2 -2\, \frac{\epsS_1-\epsS_2}{z_1-z_2} \right]\,, \qquad
{\cal B}_{\overline{\epsS},\bar{h}}^{(1)}(1,2) \equiv \bar{h} \, \left[  \bar{\partial} \overline{\epsS}_1 + \bar{\partial} \overline{\epsS}_2 -2\, \frac{\overline{\epsS}_1-\overline{\epsS}_2}{\bar{z}_1-\bar{z}_2} \right]\,.
\end{equation}
Since the general case is straightforward to deal with (see \cite{Haehl:2018izb} for details), we will simply assume the external primaries to be holomorphic, i.e., $\bar{h}=0$. Holomorphic fields only couple to the mode $\epsS$, and we can discard $\overline{\epsS}$ for ease of notation. 

\paragraph{Monodromy projection splits the soft mode propagator.}
The result for the global block computation is identical to \eqref{eq:globalBlock} with $u = z \bar{z}$ and $v = (1-z)(1-\bar{z})$, so we will not repeat it here. Instead, we observe that in two dimensions a simple split into physical and shadow contributions occurs. For instance, \eqref{eq:2dprop} separates into a sum of a (holomorphic) {\it physical part} (computing the physical block) and a {\it shadow part} (computing the shadow block):
\begin{equation}
\label{eq:2dpropSplit}
{\cal G}^E_\epsS = {\cal G}^E_{\epsS,\,\text{phys.}} + {\cal G}^E_{\epsS,\,\text{shad.}} \qquad \text{with} \qquad
\left\{ \begin{aligned}
   &{\cal G}^E_{\epsS,\,\text{phys.}} =  \frac{6}{c} \, (z_1-z_2)^2 \;  \log \left[ \mu (z_1-z_2)\right]\\
     & {\cal G}^E_{\epsS,\,\text{shad.}} =  \frac{6}{c} \, (z_1-z_2)^2 \;  \log\left[ \mu (\bar{z}_1-\bar{z}_2)\right]
     \end{aligned} \right.
\end{equation}
These two parts of the propagator are distinguished by their monodromy around $z_{12}=0$: as $z_{12} \rightarrow e^{2\pi i} z_{12}$, the physical (shadow) propagator maps to itself plus (minus) $\frac{12 \pi i}{c} \, z_{12}^2$. 

Using these two separate propagators, the two-point function of holomorphic bilinears ${\cal B}_{\epsS,h}^{(1)}$ now computes the physical and the shadow blocks separately. We find for the global identity block and shadow block between pairs of equal operators:
\begin{equation}
\label{eq:2dBSb}
\begin{split}
   G_{\Delta=2}^{(\ell=2)} &= \big{\langle} {\cal B}_{\epsS,h}^{(1)} (1,2)\,{\cal B}_{\epsS,h'}^{(1)} (3,4) \big{\rangle}_{\text{phys.}} = \frac{2hh'}{c} \; z^2 \;\, {}_2F_1(2,2,4,z) \,,\\
   G_{\tilde{\Delta}=0}^{(\ell=2)} &= \big{\langle} {\cal B}_{\epsS,h}^{(1)} (1,2)\,{\cal B}_{\epsS,h'}^{(1)} (3,4) \big{\rangle}_{\text{shad.}} = \frac{24hh'}{c}  \; \frac{\bar{z}}{z} \;\, {}_2F_1(-1,-1,-2,z) \; {}_2F_1(1,1,2,\bar{z}) \,,
\end{split}
\end{equation}
which is computed using the two parts of the propagator \eqref{eq:2dpropSplit} respectively, and we defined $z = \frac{z_{12} z_{34}}{z_{13}z_{24}}$. Note that the arbitrary scale $\mu$ has dropped out, as it must. The first line of the previous equation is, of course, just the well-known leading contribution at large $c$ to the Virasoro identity block.

\paragraph{Monodromy projection as gauging a symmetry.}
It is intriguing to see that at least in two dimensions, the separation between physical and shadow contributions already occurs at the level of the reparametrization mode propagator. We can thus implement the monodromy projection onto the physical piece at a very early stage: we simply discard the shadow part of the propagator ${\cal G}^E_\epsS$. Let us see this property at the level of the quadratic action. In $d=2$, the latter reads as
\begin{equation}
\label{eq:w2d2}
   W_2[\epsilon,\bar{\epsilon}] = \frac{ c}{24\pi} \int d^2z \;  \left[ \bar{\partial}\epsilon \; \partial^3 \epsilon  \;\; + \;\; \text{anti-holo.}\right]
\end{equation}
Following the general discussion above, consider the symmetries $\epsilon \rightarrow \epsilon+ \delta \epsilon$ of this action:
\begin{itemize}
\item The first factor, $\bar{\partial}\epsilon$, should be thought of as the source (or the shadow $\widetilde{T}$). It is invariant under 
\begin{equation}
\delta\epsilon = \Lambda_\text{h}(z)
\end{equation}
for arbitrary holomorphic functions $\Lambda_\text{h}$.
 These are chiral conformal transformations in two dimensions. 
\item The second factor, $\partial^3 \epsilon$, is proportional to the stress tensor $T[\epsilon]$ and it is invariant under 
\begin{equation}
\label{eq:2dgauge}
 \delta \epsilon = \Lambda_0(\bar{z}) + \Lambda_1(\bar{z}) \, z + \Lambda_2(\bar{z}) \, z^2 \,.
\end{equation}
These $\bar{z}$-dependent $SL(2)$ transformations were dubbed conformal redundancies above. In order to separate the physical from the shadow block, they should be understood as a gauge invariance of the theory. Gauging these is equivalent to imposing the presence of the stress tensor as a gauge invariant operator, while projecting out its shadow, as discussed above in the general case.
 \end{itemize}
 We can implement these ideas very explicitly in the present case, when we derive the propagator for $\epsilon$ via Fourier transform. For momenta $(k,\bar{k}) = (k^0+ik^1,\, k^0-ik^1)$ conjugate to the coordinates $(\bar{z},z)$, we have: 
 \begin{equation}
  \langle \epsilon(k,\bar{k}) \, \epsilon(-k,-\bar{k}) \rangle 
  = \frac{12\pi}{c} \frac{1}{\frac{k^0 + ik^1}{2}\left( \frac{k^0-ik^1}{2} \right)^3}
\end{equation}
To Fourier transform, we first integrate $k^1$ along the real line. We perform this integral via contour integration and {\it only pick up the contribution from the ``physical'' pole $k^1 = i k^0$. The ``unphysical'' pole at $k^1 = -ik^0$ is associated with the transformations \eqref{eq:2dgauge}, which we want to gauge; this means we should drop the corresponding contributions to the propagator.} This gives:
\begin{equation}
\begin{split}
    \langle \epsilon(x)\epsilon(0)\rangle_{\text{phys.}} &=
    \int_{\substack{\text{pole }\;\;\; \\ k^1 = i k^0}} \frac{d^2k}{(2\pi)^2} \; \langle \epsilon(k,\bar{k}) \, \epsilon(-k,-\bar{k}) \rangle \, e^{ik.x}
    = \frac{6}{c} \, z^2 \, \log (\mu z) + \ldots
    \end{split}
\end{equation}
where ``$\ldots$'' denotes (both finite and divergent) analytic terms.
We have thus recovered the physical propagator of \eqref{eq:2dpropSplit}. Taking into account the ``shadow pole'' $k^1=-ik^0$ in the above Fourier transform, we could have also obtained the shadow part of the propagator. This shows that (in two dimensions) the {\it monodromy projection is equivalent to gauging the transformations \eqref{eq:2dgauge}. A convenient way of doing this, at least in low dimensions, is by simply omitting the corresponding modes in the Fourier transform.}

This elucidates the monodromy projection onto the physical part of the propagator (and hence the conformal blocks etc.) at the level of the reparametrization mode, and links it to gauging a particular symmetry. While the projection onto the physical block has to be done by hand in position space, in momentum space the separation happens very naturally: the physical and shadow contributions simply come from different poles in the momentum space propagator, and imposing the gauge symmetry explicitly sets the contribution from those poles to zero. 

\vspace{10pt}
\subsection{Finite temperature}

In two dimensions it is straightforward and instructive to study thermal physics by mapping to the cylinder via 
\begin{equation}
(z,\bar{z}) = \left(\frac{\beta}{2\pi}e^{-i \frac{2\pi}{\beta} \, u},\; \frac{\beta}{2\pi}e^{i \frac{2\pi}{\beta} \, \bar{u}}\right) \,.
\end{equation}
For simplicity we set $\mu = \frac{2\pi}{\beta} = 1$. We further write $u=\tau+i \sigma$ where $\tau \in [0,2\pi)$ denotes the compact direction. The holomorphic bilinear coupling on the cylinder is simply
\begin{equation}
\label{eq:2dbilinearC}
   {\cal B}_{\epsS,h}^{(1)}(u_1,\bar{u}_1;u_2,\bar{u}_2) = h \left[ \partial \epsS_1 + \partial \epsS_2 - \frac{\epsS_1-\epsS_2}{\tan \frac{u_{12}}{2}} \right]  \,,
\end{equation}

\paragraph{Finite temperature propagator and monodromy projection.}
In order to get the physical reparametrization mode propagator, it is most convenient to follow the same strategy as in the flat space case: we start from the quadratic action to obtain the propagator in momentum space. When Fourier transforming back to position space, we omit modes that correspond to the (finite temperature version of the) gauge symmetry \eqref{eq:2dgauge}. Let us see this explicitly.

The quadratic action \eqref{eq:w2d2}, after mapping to the thermal cylinder, reads as follows:
\begin{equation}
\label{eq:W2cov2d}
W_2[\epsilon,\bar{\epsilon}]  = \frac{c}{24 \pi} \int d^2 u \; \left\{ \bar{\partial} \epsilon \; (\partial^2 + 1 ) \partial \epsilon \; + \; \text{anti-holo.}  \, 
\right\} \,.
\end{equation}
In the following we shall focus on the ``holomorphic'' first term. The ``anti-holomorphic'' second term can be treated in complete analogy.

Note that \eqref{eq:W2cov2d} looks slightly different from the action in \cite{Haehl:2018izb,Cotler:2018zff} because it is written covariantly and at this stage still contains shadow contributions. The action of \cite{Haehl:2018izb,Cotler:2018zff} is recovered by replacing $\partial \rightarrow \partial_\tau$; of course, this is a rather ad hoc prescription. We will now argue that the above action is completely equivalent to that of \cite{Haehl:2018izb,Cotler:2018zff}, as long as one gauges the appropriate symmetry associated with invariances of the stress tensor.

The action \eqref{eq:W2cov2d} for $\epsilon$ is invariant under anti-holomorphic transformations $\epsilon \rightarrow \epsilon + \Lambda_{\text{h}}(u)$ just like the flat space  action \eqref{eq:w2d2}. The additional $SL(2)$ invariance now takes the form 
\begin{equation}
\label{eq:symm2dTh}
\epsilon \rightarrow \epsilon + \delta \epsilon \,, \qquad \delta \epsilon = \Lambda_0(\bar{u}) + \Lambda_1(\bar{u}) \, e^{i u} + \Lambda_{-1}(\bar{u}) \, e^{-iu} \,.
\end{equation}
The stress tensor 
\begin{equation}
\label{eq:Teps2D}
T[\epsilon] = \frac{ c}{12}\partial(\partial^2+1)\epsilon
\end{equation}
is invariant under these transformations. In the Fourier transform below, we will therefore omit modes corresponding to these transformations in order to obtain a physical result.

The momentum space propagator corresponding to \eqref{eq:W2cov2d} is: 
\begin{equation}
\label{eq:epseps2dMom}
\begin{split} 
\langle  \epsilon(\omega_E,k) \, \epsilon(-\omega_E,-k) \rangle 
&= \frac{12\pi}{c} \frac{1}{\frac{\omega_E+ik}{2}\;\frac{\omega_E-ik}{2} \Big[ \left(\frac{\omega_E-ik}{2}\right)^2 -1\Big]  } \,,
\end{split}
\end{equation}
where $(\omega_E,k)$ are the Euclidean frequency and momentum conjugate to $(\tau,\sigma)$.
This propagator can now be Fourier transformed back to position space. This involves two steps:\footnote{  The details are analogous to calculations done in \cite{Haehl:2018izb}.} first, we need to perform a contour integral over $k$, picking up residues at the poles $k\in \{\pm i \omega_E ,\, -i(\omega_E\pm 2)\}$. Second, we need to sum over frequencies $\omega_E \in \mathbb{Z}$. Gauging the invariances of the stress tensor means that in this process we should not include modes associated with the transformations \eqref{eq:symm2dTh}. This means that in the contour integral over $k$, we only include the ``physical'' pole $k=i\omega_E$. In the subsequent sum over frequencies we omit the frequencies $\omega_E \in \{-1,0,1\}$ as they again correspond to the shadow contributions.
The contribution from the residue at $k=i\omega_E$ yields the desired result of \cite{Haehl:2018izb,Cotler:2018zff}:
{\small
\begin{equation}
\label{eq:epsprop}
\begin{split}
&\langle \epsilon(\tau,\sigma) \, \epsilon(0,0)\rangle \Big{|}_{\substack{\text{pole}\quad  \\ k =i\omega_E}}  = \frac{24}{c} \; \left[  \sin^2 \left( \frac{\tau+i\sigma}{2} \right) \log\left( 1- e^{i \, \text{sign}(\sigma)(\tau+i\sigma)} \right) -\frac{1}{4} + \frac{3}{8}\, e^{i\, \text{sign}(\sigma)(\tau+i\sigma)} \right] \equiv {\cal G}_{\epsS,\text{phys.}}^{E,\text{th.}}
\end{split}
\end{equation}
}\normalsize
The last two terms are analytic and are pure gauge in the sense that they do not contribute to physical correlation functions. We could choose to discard them. The above propagator is ${\cal G}_{\epsS,\text{phys.}}^E$ in the thermal state. It could also be obtained without any Fourier transform by conformally transforming ${\cal G}_{\epsS,\text{phys.}}^E$ and imposing translational invariance and normalizability for $|\sigma| \rightarrow \infty$ as a boundary condition. 

One can similarly check that the remaining poles $k \in \{ - i \omega_E , \, - i( \omega_E \pm 2) \}$ give contributions that reproduce precisely the thermal version of the shadow propagator ${\cal G}^E_{\epsS,\text{shad.}}$:
\begin{equation}
\label{eq:epspropShadow}
\begin{split}
\langle \epsilon(\tau,\sigma) \, \epsilon(0,0)\rangle \Big{|}_{\substack{\text{other}\\ \text{poles}}} &= \frac{24}{c}\; \sin^2 \left( \frac{\tau+i\sigma}{2} \right) \log\left( 1- e^{-i\, \text{sign}(\sigma)(\tau-i\sigma)} \right)  \equiv {\cal G}_{\epsS,\text{shad.}}^{E,\text{th.}} \,.
\end{split}
\end{equation}

The physical propagator was already written down in \cite{Haehl:2018izb,Cotler:2018zff}. Its shadow counterpart follows naturally from our analysis. The sum of these terms, ${\cal G}_\epsS^{E,\text{th.}} = {\cal G}_{\epsS,\text{phys.}}^{E,\text{th.}} + {\cal G}_{\epsS,\text{shad.}}^{E,\text{th.}}$, can be understood as the thermal version of the general result \eqref{eq:2dpropSplit} by means of a conformal transformation. 
Note that the propagators ${\cal G}_{\epsS,\text{phys.}}^{E,\text{th.}}$ and ${\cal G}_{\epsS,\text{shad.}}^{E,\text{th.}}$ are single-valued and normalizable for large spatial separations.

Finally, we remark that the physical and shadow parts are distinguished as usual by their monodromies. On the cylinder, we can focus on the monodromies around the thermal $\tau$-circle at fixed non-zero values of $\sigma$. For instance, fixing $\sigma_{12}\equiv \delta \rightarrow 0^+$, the monodromies are:
\begin{equation}
 {\cal G}^E_{\epsS,\left\{\substack{\text{phys.}\\ \text{shad.}}\right\}}  \; \stackrel{\tau \rightarrow \tau+2\pi}{\longrightarrow} \; {\cal G}^E_{\epsS,\left\{\substack{\text{phys.}\\ \text{shad.}}\right\}}  \pm 2\pi i \left[ \frac{24}{c}\,\sin^2 \left( \frac{\tau_{12}}{2}  \right)\right] \,.
\end{equation}

\paragraph{General remarks.} Note that the physical contribution to the reparametrization mode propagator (which eventually leads to Lyapunov growth of OTOCs) comes from a single pole at $k=i\omega_E$. Not coincidentally, {\it this is precisely the pole that gives rise to pole skipping in the energy-momentum tensor two-point function, which will be discussed in \S\ref{sec:EEcorr}. This leads us to conjecture that the projection onto the physical reparametrization mode propagator can be achieved in momentum space by simply only keeping the contribution from the pole that was responsible for pole skipping.}

Finally, note that this simple prescription is quite reminiscent of similar methods developed by \cite{Fitzpatrick:2011hu} in the context of conformal blocks. We can think of the projection of the momentum space integral onto a single pole in terms of multiplication of the propagator \eqref{eq:epseps2dMom} with a phase factor that imposes the gauge symmetry and removes the unphysical ``shadow poles''.
Employing a notation more similar to \cite{Cardona:2018qrt}, one has to multiply the momentum space expression for $\langle  \epsilon(\omega_E,k) \, \epsilon(-\omega_E,-k) \rangle $ with the phase function that vanishes at the unphysical ``shadow poles'' and is normalized to 1 at the physical poles. In the thermal case this function is 
\begin{equation}
\label{eq:Bdef}
B(\omega_E,k) \equiv \frac{\sin(\pi(ik-\omega_E)) \, \sin (\pi i k)}{\sin (2\pi \omega_E) \, \sin (\pi \omega_E)} \,.
\end{equation}
Multiplication with this function in momentum space achieves the projection onto physical modes only by means of removing the unphysical poles. Indeed, one easily verifies that the Fourier transform of the projected propagator, $B(\omega_E,k) \, \langle  \epsilon(\omega_E,k) \, \epsilon(-\omega_E,-k) \rangle$, is precisely \eqref{eq:epsprop} without any shadow contributions.

Having set up the building blocks of the effective theory of the reparametrization mode in two dimensions in thermal states, more interesting calculations can be performed with it. These include the ``light-light'' Virasoro block and its exponentiation \cite{Fitzpatrick:2014vua}, the ``heavy-light'' Virasoro block \cite{Fitzpatrick:2015zha} and $1/c$ corrections to it \cite{Fitzpatrick:2015dlt}, application to OTOCs \cite{Roberts:2014ifa,Fitzpatrick:2016thx}, higher-point blocks \cite{Anous:2019yku} and OTOCs \cite{Haehl:2017qfl,Haehl:2017pak}, and so on. Some of these calculations have been done for two-dimensional CFTs using the reparametrization mode formalism in \cite{Haehl:2018izb,Cotler:2018zff,Jensen:2019cmr}, so we will not repeat them here. It would, of course, be very interesting to find new applications of this formalism, as it promises to simplify calculations considerably (see \S\ref{sec:conclusion} for some suggestions). In \S\ref{sec:thermal} we will return to applications in thermal states for higher dimensions, in particular focusing on OTOCs in Rindler space.

\vspace{10pt}
\section{Application: thermal physics and OTOCs in higher dimensions}
\label{sec:thermal}

It is well known that an observer who is stationary with respect to Rindler time perceives the Minkowski vacuum state as thermal. Since the Rindler wedge of Euclidean spacetime $\mathbb{R}^d$ is conformal to a spacetime with hyperbolic spatial slices, $S^1 \times \mathbb{H}^{d-1}$, it is thus clear that a simple conformal map is sufficient for studying thermal states and quantum chaos of CFTs living on a hyperbolic space \cite{Casini:2011kv}. The temperature thus obtained is fixed in terms of the size of the $S^1$, or equivalently the curvature of the hyperbolic space; we set it to $\beta = 2\pi$ for most of this section. 

In the following, we discuss thermal out-of-time-ordered correlators in higher dimensions by various means. We start with a discussion of the OTOC based on analytically continuing the stress tensor conformal blocks (as computed by reparametrization modes). Then, we will discuss the phenomenon of pole skipping, thus establishing an even closer connection with the stress tensor two-point function.

In both cases we observe maximal Lyapunov exponent and butterfly velocity. As in two dimensions, this should be thought of as the stress tensor contribution to the OTOC, which may or may not dominate. For instance, both the maximal Lyapunov exponent and the butterfly velocity in planar ${\cal N}=4$ SYM theory receive corrections at finite couplings. This can be understood using conformal Regge theory, which takes into account contributions to the OTOC other than the stress tensor exchanges. The true Lyapunov exponent is then given by the Regge intercept associated with the leading Regge trajectory \cite{Shenker:2014cwa,Maldacena:2015waa,Mezei:2019dfv}.

\vspace{10pt}
\subsection{OTOCs from conformal blocks in higher dimensions}

In this subsection we demonstrate how to compute global stress tensor contributions to OTOCs in higher-dimensional Rindler space. The connection with our reparametrization mode formalism is as follows: we have shown in previous sections how a single reparametrization mode exchange computes global conformal blocks. We can now use these directly to study thermal physics. The basic observation is that Rindler space (a simple coordinate transform of flat Minkowski space) is conformally equivalent to a space $S^1 \times \mathbb{H}^{d-1}$. We can thus study thermal physics on hyperbolic space using just conformal symmetry arguments. We mostly repeat and expand the analysis of \cite{Perlmutter:2016pkf}, where Rindler OTOCs were computed from the global stress tensor block (note also the relevant papers \cite{Ahn:2019rnq,Mezei:2019dfv} that appeared while our work was nearing completion).

We start with the observation that the hyperbolic space metric is conformally equivalent to Rindler space:
\begin{equation}
\begin{split}
ds^2_{S^1\times\mathbb{H}^{d-1}} &= d\tau^2 + \frac{1}{\rho^2} \left( d\rho^2 + dx_\perp^2 \right) 
= \frac{1}{\rho^2} \left( \rho^2 d\tau^2  + d\rho^2 + dx_\perp^2 \right)  \equiv \Omega(\rho)^2 \, ds^2_{\text{Rindler}} 
\end{split}
\end{equation}
where $x_\perp^i \in \mathbb{R}^{d-2}$.
The conformal factor $\Omega(\rho) = \frac{1}{\rho}$ will be important when translating results back to Rindler space. 
In order to make the connection with chaos, it is most natural to work in the ``radar coordinate'' $\rho = e^{\frac{2\pi}{\beta}\sigma}$. 
The (right) Rindler wedge is in turn related to Euclidean Minkowski space $ds^2 = \eta_{\mu\nu}\, dx^\mu dx^\nu$ by the coordinate transformation 
\begin{equation}
\label{eq:Rcoords}
x^\mu = \left(  \frac{\beta}{2\pi}\,e^{\frac{2\pi}{\beta}\sigma}\,\sin\Big(\frac{2\pi}{\beta}\, \tau\Big), \,  \frac{\beta}{2\pi} \,e^{\frac{2\pi}{\beta}\sigma}\,\cos \Big(\frac{2\pi}{\beta}\tau\Big),\,x^i_\perp\right)\,,
\end{equation} 
where real Rindler time would be denoted as $t_{_\text{R}} = -i \tau$. It is therefore straightforward to study thermal physics in hyperbolic space by performing a coordinate transformation to Rindler space and then a conformal transformation. In the following we set $\beta = 2\pi$.

The quickest way to getting the thermal OTOC is by recalling our result for the global stress tensor CPW in arbitrary dimension, \eqref{eq:globalBlock}, and the physical conformal blocks associated with it. As these are simply scalar functions, all we need for the thermal setup in hyperbolic space are the Rindler expressions for the cross ratios $u$ and $v$. These follow immediately from their definition \eqref{eq:CRs} using the observation that 
\begin{equation}
x_{ij}^2  = -2 \, e^{\sigma_i+\sigma_j}\left[ \cos\tau_{ij}- \cosh \mathbf{d}(i,j)\right]\,,
\end{equation}
where $\mathbf{d}(i,j)$ is the $SO(d-2,1)$ invariant geodesic distance in the hyperbolic space $\mathbb{H}^{d-1}$ (see \eqref{eq:geodDist} for an explicit expression).

We wish to study a four-point function of pairwise equal operators inserted at (almost) coincident points. We thus set  $\mathbf{d}(1,2)=\mathbf{d}(3,4)= 0$ and $\mathbf{d}(1,3)=\mathbf{d}(1,4)=\mathbf{d}(2,3)=\mathbf{d}(2,4) \equiv \mathbf{d} = \text{const}$. We furthermore analytically continue to real time $\tau_j \rightarrow it_{_\text{R},\,j} + \delta_j$ with $t_{_\text{R},\,1}=t_{_\text{R},\,2}=t$ and $t_{_\text{R},\,3}=t_{_\text{R},\,4}=0$ where $\delta_i$ are small regulators in Euclidean time. We thus get for the cross ratios:
\begin{equation}
\begin{split}
   u &= \frac{\left[\cos\delta_{12}-1\right]\left[\cos\delta_{34}- 1\right]}{\left[\cosh(t-i \delta_{13})- \cosh\mathbf{d}\right]\left[\cosh(t-i \delta_{24})- \cosh\mathbf{d}\right]} 
   \sim e^{-2t} \; \delta_{12}^2 \delta_{34}^2 
   + \ldots\\
   v &= \frac{\left[\cosh(t-i \delta_{14})- \cosh\mathbf{d}\right]\left[\cosh(t-i \delta_{23})- \cosh\mathbf{d}\right]}{\left[\cosh(t-i \delta_{13})- \cosh\mathbf{d}\right]\left[\cosh(t-i \delta_{24})- \cosh\mathbf{d}\right]}
   \sim 1+e^{-t+\mathbf{d}} \; \delta_{12} \delta_{34} + \ldots
   \end{split}
\end{equation}
For late times $t\gg \frac{2\pi}{\beta}$, $u \approx 0$, which is precisely the short distance limit that allows us to easily distinguish the physical block from the shadow block, c.f.\ \eqref{eq:shortdist}. In terms of the complex variables $(z,\bar{z})$, we have in this limit
\begin{equation}
   z \sim -e^{-t+\mathbf{d}} \; \delta_{12} \delta_{34} \,,\qquad    \bar{z} \sim -e^{-t-\mathbf{d}} \; \delta_{12} \delta_{34} \,.
\end{equation}
This is, of course, completely analogous to the two-dimensional case with geodesic distance replacing the one spatial direction.

\begin{figure}
\definecolor{Green}{rgb}{0.25, 0.45, 0.45}
	\centering     
	\includegraphics[width=.6\textwidth]{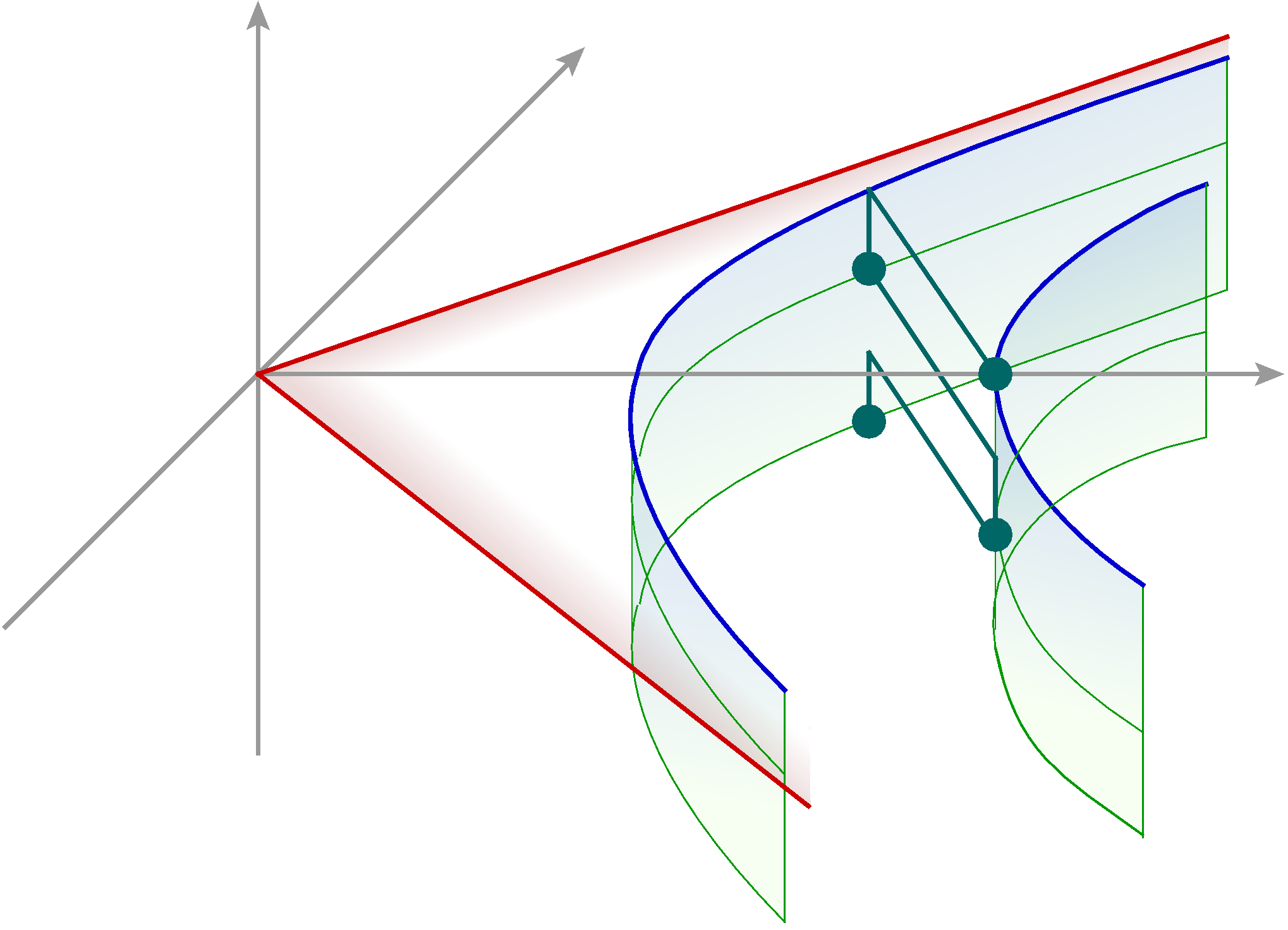}
	\put(3,111){$x^1$}
	\put(-222,197){Im$(x^0)$}
	\put(-145,185){Re$(x^0)$}
	\put(-6,176){{\color{blue}$t_{_\text{R}}$}}
	\put(-55,116){${\color{Green}W}$}
	\put(-55,76){${\color{Green}W}$}
	\put(-99,101){${\color{Green}V}$}
	\put(-99,133){${\color{Green}V}$}
	\caption{Out-of-time-order configuration in Rindler space. Inside the red shaded Rindler wedge we show two Rindler time trajectories in blue, which are at constant geodesic separation. The four operators are inserted at the green dots. We show the out-of-time-ordered configuration by including a small amount of imaginary time.}
	\label{fig:rindlerOTOC}
\end{figure}
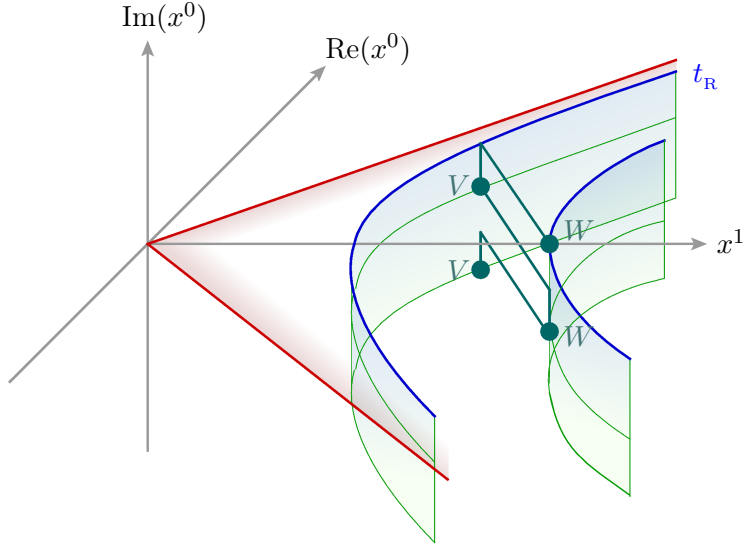

We already know that the physical block behaves in the short distance limit as \cite{Dolan:2011dv} 
\begin{equation}
\label{eq:GdAsym}
   G_{\Delta=d}^{(\ell=2)} \;\; \stackrel{u \rightarrow 0}{\sim} \;\;u^{\frac{d-2}{2}} \, (1-v)^2 \; \, {}_2F_1\left( \frac{d+2}{2} ,\, \frac{d+2}{2} ,\, d+2 ;\, 1-v \right) \,.
\end{equation}
The OTOC configuration corresponds to sending $\delta_i \rightarrow 0$ under the condition that $\delta_1>\delta_3>\delta_2>\delta_4$. As far as the multi-valued function \eqref{eq:GdAsym} is concerned, this simply means analytically continuing the cross ratio $v$ and taking it around the origin, while holding $u$ fixed. We abbreviate this operation as $(u,v) \rightarrow (u,e^{-2\pi i} v)$. Under this analytic continuation around the branch point of the conformal block, the latter picks up a monodromy, which leads to the Lyapunov behavior. The setup is illustrated in Fig.\ \ref{fig:rindlerOTOC}.

 Under the analytic continuation described above, the hypergeometric function in \eqref{eq:GdAsym} picks up a (imaginary\footnote{  The coefficient of the monodromy term is $2\pi i \, d! \, (d+1)! \, \Gamma(\frac{d}{2}+1)^{-4}$.}) monodromy $\sim(1-v)^{-d-1}$. The Euclidean block analytically continued to the Lorentzian OTOC configuration then scales as\footnote{  Notation here is somewhat imprecise with regards to spatial insertion points. We mean a configuration where $V$ operators are inserted a geodesic distance $\mathbf{d}$ away from $W$ operators.}
\begin{equation}
\begin{split}
   \frac{\langle V(t,\mathbf{d}) W(0,0) V(t,\mathbf{d}) W(0,0) \rangle_{\text{conn.}}}{\langle VV\rangle \langle WW \rangle} &\sim  G_{\Delta=d}^{(\ell=2)} (u\rightarrow 0, v \rightarrow e^{-2\pi i} v)  \\
   &\sim u^{\frac{d-2}{2}} \, (1-v)^{-d+1} \sim \frac{1}{\delta_{12}\delta_{34}} \; e^{t - (d-1) \, \mathbf{d} }   \,.
 \end{split}
\end{equation}
It follows that the Lyapunov exponent and butterfly velocity are $\lambda_L = 1$ and $v_B = \frac{1}{d-1}$, respectively, measured in units of $\frac{2\pi}{\beta}$.

\paragraph{Example in four dimensions.} As an explicit example, consider the stress tensor conformal block in $d=4$, written down in \eqref{eq:4dblocks}. Under analytic continuation to the second sheet described above, $(1-z) \rightarrow e^{-2\pi i} (1-z)$, while $\bar{z}$ is held fixed. We find: 
\begin{equation}
\begin{split}
    G_4^{(2)} &\rightarrow  -10 \left\{ \frac{\bar{z}}{z}  \frac{z^2 - 6z +6}{z-\bar{z}}  \left[ \log(1-z) - 2\pi i \right] - 6 + \frac{z}{\bar{z}}  \frac{\bar{z}^2 - 6\bar{z} +6}{\bar{z}-\bar{z}} \log(1-\bar{z})\right\} 
    \sim e^{t-3\mathbf{d}} 
    \end{split}
\end{equation}
 where we took the limit $t  - \mathbf{d} \gg \frac{\beta}{2\pi}$ in the second step. From the exponential growth we can immediately see the Lyapunov exponent is $\lambda_L = 1$ and the butterfly velocity is $v_B = \frac{1}{3}$.

\vspace{10pt}
\subsection{Chaos exponents from pole skipping in higher dimensions}
\label{sec:EEcorr}

A different motivation for the effective field theory of the soft mode revolves around the phenomenon of pole skipping, pointed out in \cite{Grozdanov:2017ajz} and further discussed in \cite{Blake:2017ris,Blake:2018leo,Grozdanov:2018kkt,Haehl:2018izb,Grozdanov:2019uhi,Blake:2019otz}. This is a phenomenon that exists in maximally chaotic theories whose Lypaunov behaviour is dominated by a single effective mode. We focus on higher-dimensional CFTs on hyperboilc (or equivalently Rindler) space. These are known to be maximally chaotic under the assumption that the stress tensor block dominates \cite{Perlmutter:2016pkf}. This analysis provides a nice consistency check with our discussion so far and generalizes \cite{Haehl:2018izb}.

The quantity we will discuss is the energy-energy conformal two-point function. While this is more complicated to compute than two-point functions of scalar operators, it is necessitated by the absence of chaos-related pole skipping in scalar two-point functions. The latter do display pole skipping, but not in a way that can be related to exponentially growing modes in an obvious way \cite{Grozdanov:2019uhi,Blake:2019otz}. While this work was nearing completion, a pole skipping analysis from the bulk point of view was done in \cite{Ahn:2019rnq}, providing results consistent with our findings.

\vspace{10pt}
\subsubsection{Thermal energy-energy correlator in CFT$_d$}

We use the previously discussed fact that thermal physics in hyperbolic space can be understood by means of a conformal transformation of a Rindler wedge. To set the stage for the discussion of pole skipping in higher dimensions, we start by working out the energy-energy two-point function on $S^1 \times \mathbb{H}^{d-1}$. We invite the reader to skip the derivation and jump to the result for the energy-energy correlator in momentum space, \eqref{eq:AnyDimRes}.

\paragraph{Position Space.}
We start by analyzing the thermal energy-energy two-point function on $S^1\times \mathbb{H}^{d-1}$ in position space. We use the embedding space formalism \cite{Costa:2014kfa} (see also Appendix \ref{sec:conventions} for more details on conventions). We embed $S^1\times \mathbb{H}^{d-1} $ in $\mathbb{R}^{1,d+1}$. The latter is a Minkowski spacetime with coordinates $X^A$, and the embedding space metric is $\eta_{AB} = \text{diag}(-1,1,\ldots,1)$.\footnote{  We denote the complete set of $\mathbb{R}^{1,d+1}$ indices as $A=(I,II,\mu)$ where the Minkowski index is $\mu = (0,m) = (0,1,i)$.} A point in $S^1\times \mathbb{H}^{d-1}$ will be parametrized as
\begin{equation}
\begin{split}
\PH^A \equiv (\PH^I,\PH^{II} ,\PH^\mu) &= \left( \frac{1+\rho^2+x_{\perp}^2}{2\rho} \,,\, \frac{1-\rho^2- x_\perp^2}{2\rho} \,,\, \sin \tau \,,\,  \cos \tau \,,\, \frac{x_\perp^i}{\rho} \right) \,,
\end{split}
\end{equation}
which lies on the projective null cone, $\PH \cdot \PH = 0$, and thus parametrizes a CFT spacetime point. 
The normalized scalar two-point function on $S^1\times \mathbb{H}^{d-1}$ for a dimension $\Delta$ operator is 
\begin{equation}
\label{eq:Gscalar}
{\cal G}_\Delta(P_1,P_2)  \equiv  \frac{1}{(-2 P_1 \cdot P_2)^\Delta}  = \frac{1}{(-2 \cos(\tau_1 - \tau_2) + 2 \cosh \mathbf{d}(1,2)  )^\Delta }\,,
\end{equation}
where $\mathbf{d}(1,2)$ is the spatial hyperbolic distance in $\mathbb{H}^{d-1}$ between the two points (see \eqref{eq:geodDist} for an explicit expression).

Since we will be interested in the chaos-related features of the energy-energy correlator, let us also record the stress-tensor two-point function (c.f., \cite{Faulkner:2015csl}):
\begin{equation} 
\label{eq:Gcorr}
\begin{split}
&{\cal G}_{\mu\nu,\rho\sigma}(P_1,P_2) \equiv \big{\langle} T_{\mu\nu}(P_1) T_{\rho\sigma}(P_2) \big{\rangle}_{S^1\times \mathbb{H}^{d-1}} \\
&\quad = 2C_T\, \mathbb{P}_{\mu\nu}^{AB}(P_1) \mathbb{P}_{\rho\sigma}^{CD}(P_2) \frac{\partial}{\partial Z^A_1} \frac{\partial}{\partial Z^B_1}\frac{\partial}{\partial Z^C_2}\frac{\partial}{\partial Z^D_2}  \left[ \frac{ \left( (P_1 \cdot P_2) (Z_1 \cdot Z_2) - (P_1 \cdot Z_2) (P_2 \cdot Z_1) \right)^2}{(-2 P_1 \cdot P_2)^{d+2}} \right] \,,
\end{split}
\end{equation}
where $Z_{1,2}$ are auxiliary vectors allowing for a compact notation. The projectors appearing in the above expression are defined as
\begin{equation}
\mathbb{P}^{AB}_{\mu\nu}(P) = \frac{\partial P^{(A}}{\partial x^\mu_{_\text{R}}} \frac{\partial P^{B)}}{\partial x^\nu_{_\text{R}}}  - \frac{1}{d} \, \eta^{AB} \eta_{CD} \,\frac{\partial P^{C}}{\partial x^\mu_{_\text{R}}} \frac{\partial P^{D}}{\partial x^\nu_{_\text{R}}} 
\end{equation}
where $x^\mu_{_\text{R}} = (\tau, \rho, x_\perp^i)$ denotes the Rindler coordinates.

We will be interested in the Fourier transform of the energy-energy correlator ${\cal G}_{00,00}(P_1,P_2)$. In order to apply known technology for performing this Fourier transform, we write the energy-energy two-point function in terms of the following auxiliary quantity:
\begin{equation}
\label{eq:Gab}
{\cal G}^{a,b}_\Delta(P_1,P_2)    \equiv \frac{1}{(-2 a \,\cos(\tau_1 - \tau_2) + 2b\,  \cosh \mathbf{d}(1,2) )^\Delta }\,.
\end{equation}
This is just the scalar two-point function \eqref{eq:Gscalar} taken slightly off-shell by the parameters $a, b$.
We can write the Euclidean energy-energy correlator in terms of this quantity as follows: 
\begin{equation}
\label{eq:G0000res}
\begin{split}
{\cal G}_{00,00}(P_1,P_2) &= 8C_T \, \bigg[ {\cal G}^{a,b}_{d+2}(P_1,P_2) + \frac{1}{4d(d+1)} \left( \frac{2(d-1)}{d} \, \partial_a \partial_b - \frac{1}{d} (\partial_a^2 + \partial_b^2 ) \right) {\cal G}^{a,b}_d(P_1,P_2) \\
&\qquad\quad  + \frac{1}{16(d-2)(d-1)d(d+1)} \, \partial_a^2 \partial_b^2 {\cal G}^{a,b}_{d-2}(P_1,P_2) \bigg]_{a,b \rightarrow 1} 
\end{split}
\end{equation}
This can be simply checked -- we provide some details in Appendix \ref{sec:deriveG0000}. This way of writing ${\cal G}_{00,00}$ will turn out to be very convenient.

\paragraph{Momentum Space.}
The next step in our calculation will be to find the Fourier transform of the energy-energy two-point function ${\cal G}_{00,00}$. The rewriting in terms of scalar propagators as in \eqref{eq:G0000res} reduces this to a simpler task: we only need to compute the Fourier transform of the slightly off-shell scalar two-point function ${\cal G}_\Delta^{a,b}$ defined in \eqref{eq:Gab}. This has been previously achieved in the on-shell case where $a=b=1$. We can therefore closely follow the calculation as presented in \cite{Ohya:2016gto}.

In order to Fourier transform, we first need to identify a suitable set of basis functions to expand in. Since we work on a spacetime $S^1 \times \mathbb{H}^{d-1}$, a basis of functions $f_{\omega,k,p_\perp}$ should be eigenfunctions of the Casimir 
\begin{equation}
\label{eq:CasimirHyper}
\Box_{S^1 \times \mathbb{H}^{d-1}} = \partial_\tau^2 + \Box_{\mathbb{H}^{d-1}} = \partial_\tau^2 + \rho^2 \left( \rho^{d-3} \partial_\rho \frac{1}{\rho^{d-3}} \partial_\rho + \Box_{\mathbb{R}^{d-2}} \right) \,. 
\end{equation}
These basis functions are characterized by the momenta $(\omega_E,k,p_\perp^i)$ conjugate to $(\tau,\rho,x_\perp^i)$. 
A suitable set of orthonormal basis functions is given by
\begin{equation}
\label{eq:fbasis}
f_{\omega,k,p_\perp}(P) = \left( \frac{4k\, \sinh(\pi k)}{\pi} \right)^{1/2} \, \rho^{\frac{d-2}{2}} \, K_{ik}( |p_\perp| \, \rho) \, e^{i (\omega_E \tau+  p_\perp\cdot\, x_\perp)} \,.
\end{equation}
These have eigenvalues $(-\omega_E^2\, , \, -k^2-\frac{(d-2)^2}{4} \,,\, - p_\perp^2)$ with respect to $\partial_\tau^2$, $\Box_{\mathbb{H}^{d-1}}$, and $\Box_{\mathbb{R}^{d-2}}$. Note that there is another set of solutions of the same form, but with  $J_{ik}(i|p_\perp| \, \rho)$ instead of $K_{ik}(|p_\perp| \, \rho)$. These are not normalizable, so we discard them. However, below we will discuss imaginary values for the momentum $k$, in which case the regular solutions are precisely these Bessel-$J$ functions.

With respect to this basis of orthonormal functions, we find that the momentum space expression for ${\cal G}_\Delta^{a,b}$ is given by the following: 
\begin{equation}
\label{eq:GDeltaMom}
\begin{split}
{\cal G}_\Delta^{a,b}(P_1,P_2) &= \frac{1}{(2\pi)^d} \sum_{\omega_E} \int_0^\infty dk \int d^{d-2} p_\perp \;\; {\cal G}_\Delta^{a,b}(\omega_E,k) \, f_{\omega_E,k,p_\perp}(P_1) \, f^*_{\omega_E,k,p_\perp}(P_2)  \,,
\end{split}
\end{equation}
where 
\begin{equation}
\label{eq:GDeltaMom2}
\begin{split}
{\cal G}_\Delta^{a,b}(\omega_E, k) &=  \frac{\pi^{d/2}}{\Gamma(\Delta)} \frac{a^{\omega_E}}{b^{\Delta+ \omega_E}} \frac{ \Gamma(\alpha)\Gamma(\alpha^*)}{\Gamma(\omega_E+1)} \; {}_2F_1\left( \alpha, \, \alpha^*, \, \omega_E+1 ; \, \frac{a^2}{b^2} \right)_\text{reg.}
\end{split}
\end{equation}
with $\alpha \equiv \frac{1}{2} (\Delta - \frac{d-2}{2} + ik + \omega_E )$. The subscript ``reg.'' refers to the fact that the expression needs to be regularized in the limit as $a,b \rightarrow 1$ whence the hypergeometric function is naively divergent. We will describe this regularization below. The derivation of \eqref{eq:GDeltaMom2} is analogous to the case of on-shell scalar two-point functions in \cite{Ohya:2016gto}. We review some of the details in Appendix \ref{sec:ohyaCalc}.

By combining \eqref{eq:GDeltaMom2} with \eqref{eq:G0000res} (in momentum space) we can now immediately infer the desired correlation function. Taking derivatives with respect to $a$ and $b$ as in \eqref{eq:GDeltaMom2}, we find for the {\it energy-energy two-point function on} $S^1 \times \mathbb{H}^{d-1}$:
\begin{equation}
\label{eq:AnyDimRes}
\boxed{
	\begin{split}
	&{\cal G}_{00,00}(\omega_E,k)  \\
	&\quad = \frac{C_T\, \pi^{d/2} \,(d-2)\, \Gamma(-\frac{d}{2})}{8 \, (d+1) \,\Gamma(d) } \,  \left( k^2 + \left(\frac{d}{2}\right)^2 \right) \left( k^2 + \left( \frac{d-2}{2}\right)^2 \right) \,  \frac{\Gamma\left[ \frac{1}{2} \left(\omega_E \pm ik + \frac{d-2}{2} \right) \right] }{\Gamma\left[ \frac{1}{2} \left( \omega_E\pm ik -\frac{d-6}{2}  \right) \right] } \Bigg{|}_\text{reg.}
	\end{split}
}
\end{equation}
Here and henceforth, any double sign ``$\pm$'' inside a function's argument means that one should multiply by all possible signs; for example, $\Gamma(x\pm y) \equiv \Gamma(x+y) \Gamma(x-y)$. Further, the expression is divergent for even dimensions due to the factor $\Gamma(-\frac{d}{2})$. This can be dealt with by performing dimensional regularization and keeping the finite term.
We discuss this regularization in some more detail in Appendix \ref{sec:EvenDimReg}.

\vspace{10pt}
\subsubsection{Pole skipping} 

Given the momentum space expression for the energy-energy correlator, we now wish to discuss the pole skipping phenomenon as described in \cite{Blake:2017ris}. There, it was argued that it is interesting to consider the analytic continuation of the energy-energy correlator to complex frequencies and momenta. The latter has lines of poles, which are however lifted at certain special points. These special points allow one to read off the Lyapunov exponent and butterfly velocity in case of chaotic theories. 

Our discussion below will generalize the study of pole skipping in two-dimensional CFTs \cite{Haehl:2018izb,Das:2019tga}. As we shall see, a very similar pole skipping is observed in arbitrary dimensions and it is sensibly related to the Lyapunov exponent and butterfly velocity characterizing the OTOC. We discuss even and odd dimensions separately.

\paragraph{Odd dimensions.} For simplicity, we first focus on the odd-dimensional case, where we do not need any further regularization of \eqref{eq:AnyDimRes}.

First note that the pole structure of \eqref{eq:AnyDimRes} comes from the following singular part:
\begin{equation}
\label{eq:Gsing}
{\cal G}_{00,00}(\omega_E,k) \Big{|}_\text{singular} \sim  \Gamma\left[ \frac{1}{2} \left( \omega_E + ik +  \frac{d-2}{2}\right) \right] \,  \Gamma\left[ \frac{1}{2} \left( \omega_E - ik + \frac{d-2}{2}  \right) \right]
\end{equation}
The two $\Gamma$-functions in the denominator manifestly don't have any zeros (for finite values of the arguments) and hence do not contribute any poles to the two-point function. 
This singular piece \eqref{eq:Gsing} has...
\begin{equation}
\label{eq:poles1}
\text{...lines of simple poles at: } \quad k =  \pm i \left[    \frac{d-2}{2}  + 2n 
+  \omega_E\right] \qquad\quad   (n=0,1,2,\ldots)
\end{equation}
We will refer to these as the {\it pole lines}. 

In addition, the propagator ${\cal G}_{00,00}(\omega_E,k)$ has {\it zero lines}. These are of two types: there are four vertical lines due to the vanishing of the prefactor $(k^2+(\frac{d}{2})^2)(k^2+(\frac{d-2}{2})^2)$, leading to...
\begin{equation}
\label{eq:zeros1}
\text{...lines of simple zeros at:} \qquad\quad k =  \pm i  \, \frac{d}{2}  \quad \text{and} \quad  k =  \pm i \, \frac{d-2}{2} \,.
\end{equation}
Furthermore, the denominator of \eqref{eq:AnyDimRes} has simple poles, leading at the level of the correlation function to...
\begin{equation}
\label{eq:zeros2}
\text{...lines of simple zeros at: } \quad k =  \pm i \left[   - \frac{d-6}{2}  + 2m 
+  \omega_E\right] \qquad\quad   (m=0,1,2,\ldots)
\end{equation}

\begin{figure}
	\centering     
	\includegraphics[width=.45\textwidth]{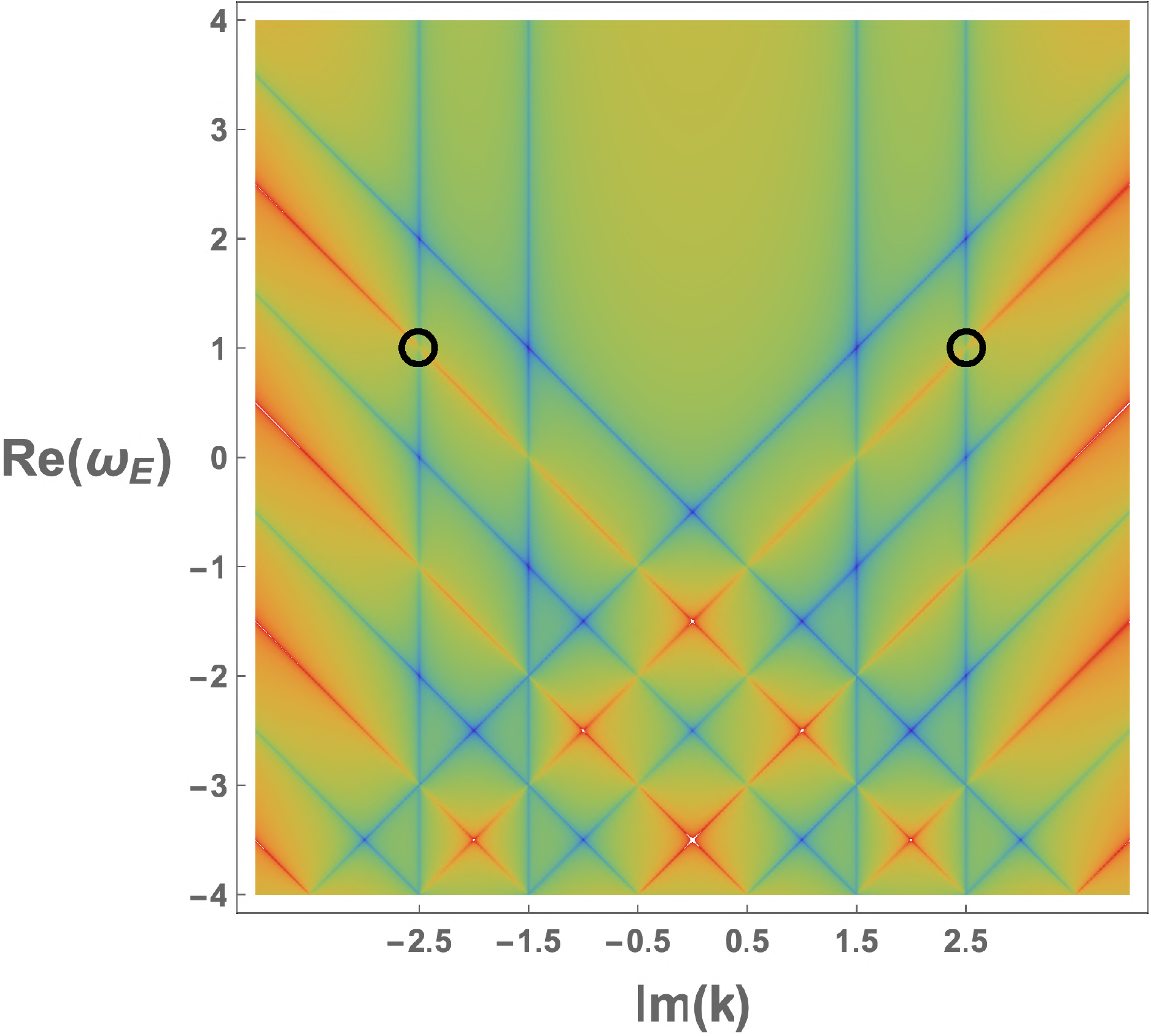}$\qquad$
	\includegraphics[width=.45\textwidth]{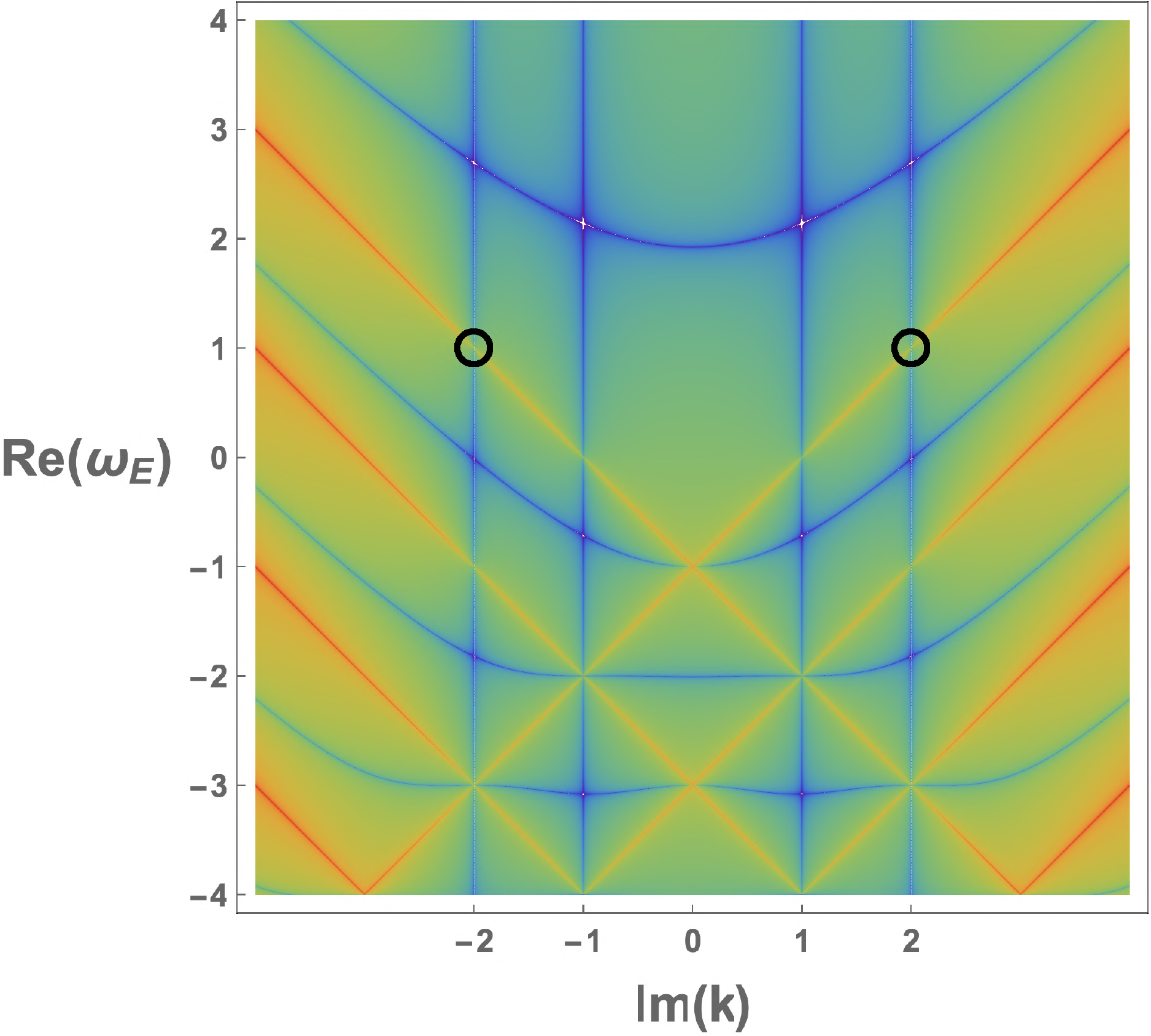}
	\caption{Plots of $\log | {\cal G}_{00,00}(\omega_E,k) |$ in $d=5$ (left) and $d=4$ (right). The ``warm'' lines (orange/red) correspond to poles. The ``cold'' lines (blue) correspond to zeros. The four vertical zero lines are due to the polynomial prefactors in \eqref{eq:AnyDimRes} and \eqref{eq:G0000even}. Pole skipping occurs when zero lines and pole lines intersect. In order to make a connection with exponentially growing modes in quantum chaos, one should focus on the upper half plane, where pole skipping is observed at precisely two locations (black circles): $(\omega_E , k)_\text{skip} =  (1, \pm i \frac{d}{2})$. Note that in the even dimensional case (right figure) the precise shape of the zero lines (but not the pole skipping locations) depend on ambiguous contact terms.}
	\label{fig:G0000odd}
\end{figure}

The situation is best understood by drawing a density plot of ${\cal G}_{00,00}(\omega_E,k)$, see the left panel of Fig.\ \ref{fig:G0000odd}. This illustrates clearly the structure of zeros and poles (for the example of $d=5$).

We are interested in the pole skipping phenomenon, i.e., the locations where pole lines and zero lines intersect and the pole is lifted. As we can easily see, this happens infinitely often. However, let us focus on the upper half plane, $\text{Re}(\omega_E) > 0$, where poles correspond to exponentially growing modes relevant to chaos. In the upper half plane the zero lines \eqref{eq:zeros1} never intersect with pole lines \eqref{eq:poles1} and pole skipping is solely due to the vertical zero lines \eqref{eq:zeros2}. Furthermore, only the line $k=\pm i \frac{d}{2}$ intersects with the pole lines indexed by $n=1$. In order to get exponential growth, our conventions will single out the following pole skipping locations as the relevant one:
\begin{equation}
\label{eq:skip1}
(\omega_E , k)_\text{skip} =  \Big( 1, \; \pm  i \,\frac{d}{2} \Big)\,.
\end{equation} 
For later reference, we also record the other three locations where $n=1$ pole lines are skipped due to intersection with the vertical zero lines:
\begin{equation}
\label{eq:skipS}
  (\omega_E , k)_\text{skip}^{(s)} =  \Big(0 ,\;  \pm i \, \frac{d-2}{2} \Big)  \,.
\end{equation}
The meaning of the superscript ${}^{(s)}$ is supposed to indicate that these pole skipping locations are related to ``shadow'' contributions to the soft mode propagator.

We will demonstrate below how to determine the Lyapunov exponent of chaotic CFTs from \eqref{eq:skip1}. It is crucial to stress that this particular pole skipping occurs for complex frequency in the {\it upper} half plane, thus leading to exponentially growing modes. This is a feature of the stress tensor correlator, which is absent in discussions of other two-point functions that also display pole skipping, but only in the lower half plane and not related to quantum chaos in any immediate way (c.f.\ \cite{Grozdanov:2019uhi,Blake:2019otz}).

\paragraph{Even dimensions.}
In even dimensions, the regularization of the expression \eqref{eq:AnyDimRes} leads to a slightly more complicated result. We offer the full expression in Appendix \ref{sec:EvenDimReg}. For now, let us simply quote the result \eqref{eq:EvenDimRes} for the special case of $d=4$:
\begin{equation}
\label{eq:G0000even}
\begin{split}
{\cal G}_{00,00}(\omega_E,k)_{\text{finite}}^{(d =4)} &= - \frac{C_T\, \pi^2}{240} \, (k^2 + 4)(k^2 + 1)  \,  \bigg\{  \psi\left( \frac{\omega_E+ik + 1}{2}\right) + \psi\left( \frac{\omega_E-ik +1 }{2}\right)    \bigg\} 
\end{split} 
\end{equation}
A section of this function is plotted in the right panel of Fig.\ \ref{fig:G0000odd}.

As is clear from inspection, the lines of poles are precisely the same as in odd dimensions, \eqref{eq:poles1}. Also, the vertical lines of zeros due to the polynomial prefactor are identical to \eqref{eq:zeros1}. Only the additional lines of zeros (the analog of \eqref{eq:zeros2}) are more complicated and in fact depend on contact terms. As far as pole skipping is concerned, we will mostly be interested in the intersection of pole lines and vertical zero lines, which are identical to the odd dimensional case and independent of the regularization and contact terms; unsurprisingly, they lead to pole skipping at the same location as in odd dimensions, \eqref{eq:skip1}.

\paragraph{Lyapunov exponent and butterfly velocity from pole skipping.}

Let us focus on the $n=0$ trajectory in the upper half plane (for any $d\geq 3$), where the pole skipping happens at $(\omega_E, k)_\text{skip} = ( 1,  \,\pm i \frac{d}{2})$. We wish to interpret these locations as characterizing the Lyapunov exponent and butterfly velocity in Rindler space. In order to interpret this, we will now give two arguments -- one abstract and one very concrete. We will set the transverse momenta $p_\perp^i=0$ for simplicity.

As an abstract interpretation of the pole skipping on the $n=0$ trajectory, note that the pole skipping at $(\omega_E, k)_\text{skip} = ( 1, \pm\, i \frac{d}{2})$ corresponds to the following eigenvalues of the Casimir \eqref{eq:CasimirHyper} on $S^1 \times \mathbb{H}^{d-1}$:
\begin{equation}
\left(\omega_E^2 , \;  -k^2 - \frac{(d-2)^2}{4} \right)_\text{skip}  = (1,\, d-1) \,,
\end{equation}
where we neglect the transverse directions. As in the flat space case, we interpret the first eigenvalue as saying that there is an imaginary frequency mode, corresponding to exponential growth with Lyapunov exponent $1$ in units of $\frac{2\pi}{\beta}$. Similarly, we interpret the second eigenvalue as the inverse of the butterfly velocity in the same units. That is, $v_B = \frac{1}{d-1}$ in accordance with \cite{Perlmutter:2016pkf}.\footnote{  As a side remark, note that the other vertical zero line, leading to pole skipping at $(\omega_E,k) = (0, \, \pm i \frac{d-2}{2})$ has eigenvalues $\big(\omega_E^2 , \;  -k^2 - \frac{(d-2)^2}{4} \big)_\text{skip} = (0,0)$. This seems unrelated to Lyapunov growth.}

 In order to give a more concrete interpretation, we can evaluate the basic ``Fourier'' modes \eqref{eq:fbasis} at the pole skipping location in a long-time (Lyapunov) and short-distance (flat space) limit. Obviously, the time dependence of \eqref{eq:fbasis} is just a plane wave, which evaluated at $\omega_E = \pm i$ gives exponential modes $e^{\mp \tau}$ associated with a Lyapunov exponent $\lambda_L =1$. In order to extract the butterfly velocity, we study the $u$-dependence of \eqref{eq:fbasis} at the pole skipping location for spatially nearby points ($\rho \rightarrow 0$). One subtlety is that the Bessel function $K_{ik}$ evaluated at $k=i\, \frac{d}{2}$ is not regular at short distances. In order to get a sensible interpretation, we replace it by the regular solution to the Casimir on $S^1 \times \mathbb{H}^{d-1}$, eq.\ \eqref{eq:CasimirHyper}, which takes the form of \eqref{eq:fbasis} with Bessel-$J$ functions instead of Bessel-$K$. At the pole skipping location these hyperbolic eigenfunctions behave as 
 \begin{equation}
   \rho^{\frac{d-2}{2}} J_{i k} (i|p_\perp| \, \rho)  \Big|_{(\omega_E,k) \rightarrow (1,-i\frac{d}{2})} 
      \stackrel{\rho \rightarrow 0}{\sim} \;\;\rho^{d-1}  \,.
 \end{equation}
Writing $\rho=e^\sigma$, this behavior clearly corresponds to a butterfly velocity $v_B = \frac{1}{d-1}$ as expected for chaos in Rindler space.

\vspace{10pt}
\section{Conclusions and outlook}
\label{sec:conclusion}

\paragraph{Summary.}
Motivated by recent studies of reparametrization modes in the low energy SYK model and two-dimensional CFTs, we set out to provide a more general understanding of the dynamics of such modes in CFTs of arbitrary dimension. We have established a variety of connections between reparametrization modes, the conformal anomaly, the shadow operator formalism for conserved currents, and OPE block techniques in kinematic space. These connections explain in detail how the reparametrization modes are related to the physics of stress tensor contributions to conformal blocks and provide an interesting new perspective on all these topics. 

The simplest example for a physical computation is the exchange of a single reparametrization mode in a four-point function. At large central charge this is indeed the dominant contribution. It computes the single-stress tensor exchange in the global conformal block. Exchanges of multiple reparametrization modes are expected to compute multi-stress tensor exchanges. In $d=2$ these compute the stress tensor contribution to the Virasoro identity block; in the ``light-light'' regime, the dominant diagrams are ladder diagrams.

As an immediate application, we also studied CFTs in thermal states by exploiting the equivalence of physics in the Rindler wedge and thermal physics of CFTs living on a hyperbolic space. By investigating pole skipping in the thermal energy-energy two-point function in arbitrary dimensions, we confirmed the expected behavior of out-of-time-ordered correlation functions and derived the chaos exponents for maximally chaotic CFTs on hyperbolic space.

\paragraph{Questions and future directions.}
Some interesting opportunities for future research are now available. While we provided a quadratic action for the reparametrization mode, we have not yet found a way to complete it nonlinearly. In one dimension, the nonlinear completion is the Schwarzian action \cite{Kitaev:2015aa,Maldacena:2016hyu}; in two dimensions the nonlinear completion is the Alekseev-Shatashvili action \cite{Alekseev:1988ce,Cotler:2018zff}. We wonder if it might be possible to obtain a nonlinear action in higher (even) dimensions by performing a coadjoint orbit quantization of kinematic space \eqref{eq:kinspace}. It was already shown in \cite{Penna:2018xqq} that higher-dimensional kinematic space is indeed a coadjoint orbit of the conformal group. It would be interesting to explore this further in light of the new perspective provided by the present paper.\footnote{  See also \cite{Callebaut:2018xfu,Callebaut:2018nlq} for other relations of kinematic space with theories of reparametrizations in $d=2$.}

So far, all our detailed understanding of chaotic theories in terms of few effective degrees of freedoms (reparametrization modes) describes theories that saturate the chaos bound of \cite{Maldacena:2015waa}. It will be interesting to understand how the reparametrization mode theory should be modified such as to capture corrections to the maximal Lyapunov exponent. It is clear that the pole skipping phenomenon is special to maximally chaotic theories (for instance, as we have shown, it occurs in any conformal field theory, irrespective of integrability, chaoticity etc.). It would be interesting if there exists a suitable generalization of the pole skipping phenomenon which provides an effective field theory description of modes other than just stress tensor contributions and is thus capable of capturing corrections to maximal chaos.

It will be interesting to apply the theory of reparametrization modes to study stress tensor conformal blocks at large central charge. These are particularly interesting in the context of holography, where they describe the leading universal physics of graviton exchanges in the bulk. Various known results have already been reproduced, illustrating the utility and simplicity of the formalism. It should now be used to find novel results -- for example, one can imagine studying conformal blocks in new kinematic regimes (e.g., the late time regime \cite{Kraus:2018zrn}, the lightcone regime \cite{Kusuki:2019gjs}, etc.), or higher-point conformal blocks in various channels \cite{Rosenhaus:2018zqn,Anous:2019yku,us:wip}. Furthermore, not much is known about the higher-dimensional conformal blocks beyond the global single-stress tensor exchanges (see however \cite{Fitzpatrick:2019zqz,Fitzpatrick:2019efk,Kulaxizi:2019tkd}).

Some technicalities about the formulation of the shadow operator formulation in terms of the reparametrization mode deserve better understanding. For instance, much of our formulation is restricted to an even number of spacetime dimensions: the quadratic action for the reparametrization mode only exists due to the conformal anomaly in even dimensions. We anticipate that the non-existence of a simple theory of reparametrizations in odd dimensions might be related to various other complications in that case. For instance, even global conformal blocks are very complicated in odd dimensions \cite{Dolan:2011dv}. On the other hand, the pole skipping phenomenon discussed in \S\ref{sec:EEcorr} does occur both in even and odd dimensions.\footnote{  Note, however, that the technical details are quite different in the two cases.} It would be good to understand what this means for the existence or non-existence of a simple theory of reparametrizations in odd dimensions.

\vspace{10pt}
\acknowledgments

We would like to thank Tarek Anous, Sean Cooper, Jordan Cotler, Jan de Boer, Daniel Grumiller, Eliot Hijano, Kristan Jensen, Alex May, Masamichi Miyaji, Dominik Neuenfeld, David Ramirez, Mukund Rangamani, Douglas Stanford, James Sully, Chris Waddell, and David Wakeham for enlightening discussions. MR thanks the Amsterdam String Workshop for hospitality, and many of its participants for interesting discussions. FH appreciates support and hospitality by the KITP Santa Barbara and the Aspen Center for Physics. Research at the KITP is supported in part by the National Science Foundation under Grant No.\ PHY-1748958. The Aspen Center for physics is supported by National Science Foundation grant PHY-1607611. The work of FH was also supported by the Simons Collaboration {\it It From Qubit}, as well as the US Department of Energy under grant DE-SC0009988.

\appendix 

\vspace{10pt}
\section{Conventions and useful formulae}
\label{sec:conventions}

\paragraph{Conventions in $d=2$.}
For reference, we collect some of our conventions for calculations in two-dimensional CFTs. On the Euclidean plane $x^\mu = (x^0,x^1)$, we define $z=x^0 + i x^1$ and $\bar{z} = z^*$. After mapping to the cylinder via $z = e^{-iu}$, we similarly write $u=\tau + i \sigma$. These conventions yield: 
\begin{equation}
\begin{split}
   \text{metric:} \qquad & ds^2 = dz \, d\bar{z}  = (dx^0)^2 + (dx^1)^2 \\
   &\quad\;\,  = e^{i(\bar{u}-u)}\, du \, d\bar{u} = e^{2\sigma} \, \left( d\tau^2 + d\sigma^2 \right) \,,\\
   \text{derivatives:} \qquad 
   & \partial_u \equiv \frac{1}{2}(\partial_\tau - i \partial_\sigma) \,,\quad \bar{\partial}_{\bar{u}} \equiv \frac{1}{2}(\partial_\tau + i \partial_\sigma) \,,\\
   \text{integral measure:} \qquad & \int d^2z  \equiv  \frac{i}{2} \int dz d\bar{z} =  \int d^2 x   \,,\qquad 
   \int d^2u \equiv \int d\tau d\sigma \,.
\end{split}
\end{equation}
We define the holomorphic stress tensor on the plane as $T \equiv -2\pi T_{zz}$, and similarly $\overline{T} \equiv -2\pi T_{\bar z \bar z}$.

\paragraph{Conventions in $d \geq 2$.}
In higher dimensions, we often use the symmetric-traceless projector and the inversion tensor:
\begin{equation}
\label{eq:TTproj}
\begin{split}
   \mathbb{P}\,{}^{\mu\nu}_{\rho\sigma} &\equiv \frac{1}{2} \left(  \delta^\mu_\rho \delta^\nu_\sigma + \delta^\nu_\rho \delta^\mu_\sigma \right) - \frac{1}{d}  \, \eta_{\rho\sigma} \eta^{\mu\nu} \,,\qquad\;\;\;
   I^{\mu\nu}(x) \equiv \eta^{\mu\nu} - 2 \, \frac{x^\mu x^\nu}{x^2} \,.
   \end{split}
\end{equation}
The inversion tensor satisfies the following useful relations: 
\begin{equation}
\label{eq:Irelations}
\begin{split}
   \frac{1}{x^2} \, I_\nu^\mu(x) &=   \partial_\nu \left( \frac{x^\mu}{x^2} \right)  = \partial_\nu\partial^\mu \log |x|  \,,\qquad\;\;
   \partial_{\mu} I_{\nu}^\rho = - \frac{2}{x^2} \left[ I_{\mu}^\rho \; x_{\nu}+  x^\rho  \, \eta_{\mu\nu}\right] \,.
\end{split}
\end{equation}

\paragraph{Conventions for embedding space.} 
A general point in embedding space $\mathbb{R}^{1,d+1}$ is denoted as $X^A$, where $A = (I,I\!I,\mu) = (I,I\!I,0,m)$. The metric is $\eta_{AB} = \text{diag}(-1,1,\ldots,1)$. Let us consider points $P^A$ that lie on the (upper) projective null cone characterized by $P\cdot P = 0$ and $P^I > 0$. The CFT spacetime consists of points on the null cone subject to the projective identification $P^A \equiv \lambda P^A$. A point $P^A$ in the CFT can be represented in different useful ways. We mainly use the following representations:
\begin{equation}
\label{eq:coordsEmb}
\begin{split}
  \text{Poincar\'{e} section:} \qquad \PP^A &= \left( \frac{1+x^2}{2} \,, \, \frac{1-x^2}{2} \,,\, x^\mu \right)  \qquad x^\mu \in \mathbb{R}^d\\
  \text{Rindler section:} \qquad \PR^A &= \left( \frac{1+\rho^2+x_{\perp}^2}{2} \,,\, \frac{1-\rho^2- x_\perp^2}{2} \,,\, \rho \, \sin \tau \,,\, \rho \, \cos \tau \,,\, x_\perp^i \right) \,,\\
  \text{Hyperbolic space:} \qquad \PH^A &= \frac{1}{\rho} \; \PR^A\,, \qquad \rho >0 \,,\; x^i_\perp \in \mathbb{R}^{d-1}\,.
\end{split}
\end{equation}
The Poincar\'{e} section is appropriate for studying zero-temperature CFTs on $d$-dimensional Minkowski spacetime (or simply Euclidean $\mathbb{R}^d$ as above). For instance, in Poincar\'{e} coordinates we have $(-2 P_{1} \cdot P_{2}) = (x_1-x_2)^2$. 

On the other hand, the Rindler coordinates correspond to studying CFTs, which look thermal with respect to Rindler time $\tau$. The temperature is fixed as $T=2\pi$. The hyperbolic section is conformal to Rindler space, and the conformal factor can simply be read off from the third line of \eqref{eq:coordsEmb}.
In the hyperbolic (or Rindler) section it is useful to parametrize the hyperbolic space $\mathbb{H}^{d-1}$ by itself (without the time direction) in terms of
\begin{equation}
  Y^A = \left( \frac{1+\rho^2 + x_\perp^2}{2\rho} ,\, \frac{1-\rho^2-x_\perp^2}{2\rho} ,\,0,\,0,\, \frac{x^i_\perp}{\rho} \right)\,.
\end{equation}
The $SO(d-2,1)$ invariant geodesic distance in $\mathbb{H}^{d-1}$ is 
\begin{equation}
\label{eq:geodDist}
  \mathbf{d}(1,2) = \cosh^{-1} (-Y_1 \cdot Y_2) = \cosh^{-1} \left[ \frac{\rho_1^2 + \rho_2^2 + x_{\perp_{12}}^2}{2\, \rho_1 \rho_2} \right]\,.
\end{equation}
We sometimes change to a radial hyperbolic coordinate, $\rho = e^\sigma$.

\paragraph{Conformal integrals.} We frequently write integrals over all of Euclidean spacetime. These are typically determined by conformal symmetry, and evaluating them in practice is most conveniently done in embedding space, see e.g.\ \cite{SimmonsDuffin:2012uy} for examples, some of which we review now. Embedding space integrals in higher dimensions should be understood as follows:
\begin{equation}
   \int d^dP \; f(P) \equiv  \, \int_{P^{I} > 0} \frac{d^{d+2} P}{\text{Vol GL}(1,\mathbb{R})^+} \, \delta(P^2) \, f(P) \,.
\end{equation}
where division by the volume of the gauge group of planar boosts associated with identification of points under rescaling ($P^A \equiv \lambda P^A$) renders the result finite. For example, in the coordinates introduced above, these definitions reduce to:
\begin{equation}
\begin{split}
  \int d^dP  \; f(P) \equiv \int d^dx \; f(\PP) \equiv \int_0^{2\pi} d\tau \int_0^\infty \frac{d\rho}{\rho^{d-1}} \int d^{d-2} x_\perp \; \Omega(\PH)^{-\Delta}\, f(\PH)\,,
\end{split}
\end{equation}
where $\Delta$ is the conformal weight of $f(P)$.

An important example of an embedding space integral, that can be used to check various normalizations in this paper, is the following:
\begin{equation}
\label{eq:doubleint}
  \iint  \frac{d^dP_1 \, d^dP_2}{(-2 X_1\cdot P_1)^{\Delta} (-2X_2 \cdot P_2)^{d-\Delta} (-2 P_1 \cdot P_2)^\Delta } = \frac{\pi^d \Gamma(\Delta- \frac{d}{2}) \Gamma(\frac{d}{2}-\Delta)}{\Gamma(\Delta) \Gamma(d-\Delta)}  \frac{1}{(-2 X_1 \cdot X_2)^\Delta} \,,
\end{equation}
and similarly for conformal integrals involving spinning propagators, where the normalization on the right hand side gets replaced by $\pi^d (k_{\Delta,\ell} \,k_{d-\Delta,\ell})^{-1}$, c.f., \eqref{eq:kNorm}. 

\paragraph{Shadows and projectors.} We can use \eqref{eq:doubleint} to check various normalizations in \S\ref{sec:shadows}. As a first application, we determine the normalization of two-point functions of shadow operators in terms of the normalization of their physical partners. Define the two-point function of scalar operators ${\cal O}$ as $\langle {\cal O}(P_1) {\cal O}(P_2) \rangle  = \frac{C_{\cal O}}{(-2P_1\cdot P_2)^\Delta} = \frac{C_{\cal O}}{(x_1 - x_2)^{2\Delta}}$. Using the definition of the shadow operator \eqref{eq:QOhigherd}, it follows immediately from \eqref{eq:doubleint} that 
\begin{equation}
\langle \widetilde{\cal O}(P_1) \widetilde{\cal O}(P_2) \rangle  = \frac{C_{\widetilde{\cal O}}}{(-2P_1\cdot P_2)^\Delta}  \qquad \text{with}\qquad  C_{\widetilde{O}} = C_{\cal O} \, \frac{k_{\Delta,0}}{k_{d-\Delta,0}} \,,
\end{equation}
and similarly for spinning operators.

Let us check more features of the definition of the shadow operator, \eqref{eq:QOhigherd}. In order to show that $\doublewidetilde{\cal O} = {\cal O}$, one can, for example, verify that $\langle \doublewidetilde{\cal O}(x) {\cal O}(y) \rangle = \langle {\cal O}(x) {\cal O}(y) \rangle$ using \eqref{eq:doubleint}.
In general we define the projectors onto the conformal family associated with an operator ${\cal O}$ and its shadow as:
\begin{equation}
\begin{split}
   |{\cal O}| &\equiv \frac{1}{C_{\cal O}}\frac{k_{d-\Delta,\ell}}{\pi^{d/2}}  \int d^d \xi \; {\cal O}^{\mu \cdots}(\xi) |0\rangle \langle 0| \widetilde{\cal O}_{\mu \cdots }(\xi) \,,\\
   |\widetilde{\cal O}| & \equiv \frac{1}{C_{\widetilde{\cal O}}}\frac{k_{\Delta,\ell}}{\pi^{d/2}}  \int d^d \xi \; \widetilde{\cal O}^{\mu \cdots}(\xi) |0\rangle \langle 0| {\cal O}_{\mu\cdots}(\xi)
\end{split}
\end{equation} 
where $C_{\widetilde{\cal O}} \equiv C_{\cal O} \, \frac{k_{\Delta,\ell}}{k_{d-\Delta,\ell}}$ as above. This ensures that $| {\cal O} | = |\widetilde{\cal O}|$ as is obvious from the definition \eqref{eq:QOhigherd}. One can furthermore check that $|{\cal O}|$ leaves correlators involving the operator ${\cal O}$ invariant. For instance, one easily checks that $\langle {\cal O}(x) |{\cal O}| {\cal O}(y) \rangle = \langle {\cal O}(x) {\cal O}(y) \rangle$, which is again a direct consequence of the integral \eqref{eq:doubleint}. Similarly, the fact that $|{\cal O}|^2 = |{\cal O}|$ is another simple application of \eqref{eq:doubleint}. For more general statements, see \cite{SimmonsDuffin:2012uy}.

\vspace{10pt}

\section{Exponentiation of the light-light Virasoro block}
\label{sec:LLLL}
As a simple illustrative example of the reparametrization mode formalism, we briefly review the exponentiation of the light-light conformal block in $d=2$ as presented in \cite{Cotler:2018zff}. In the ``light-light'' regime we take the external holomorphic operators $V$ and $W$ to have dimensions $h_V = h_W \equiv h \sim \sqrt{c}$. The leading reparametrization mode exchanges in this regime correspond to diagrams of order $(h^2/c)^n \sim {\cal O}(1)$. These diagrams are the ladder diagrams of the following form: 
\vspace{-.2cm}
\begin{equation*}
\label{eq:LLLLdiag}
\begin{split}
 {\cal V}_0 \big{|}_{{\cal O}(1)} \quad &=  \quad 1 \quad 
\begin{gathered}
\includegraphics[width=.4\textwidth]{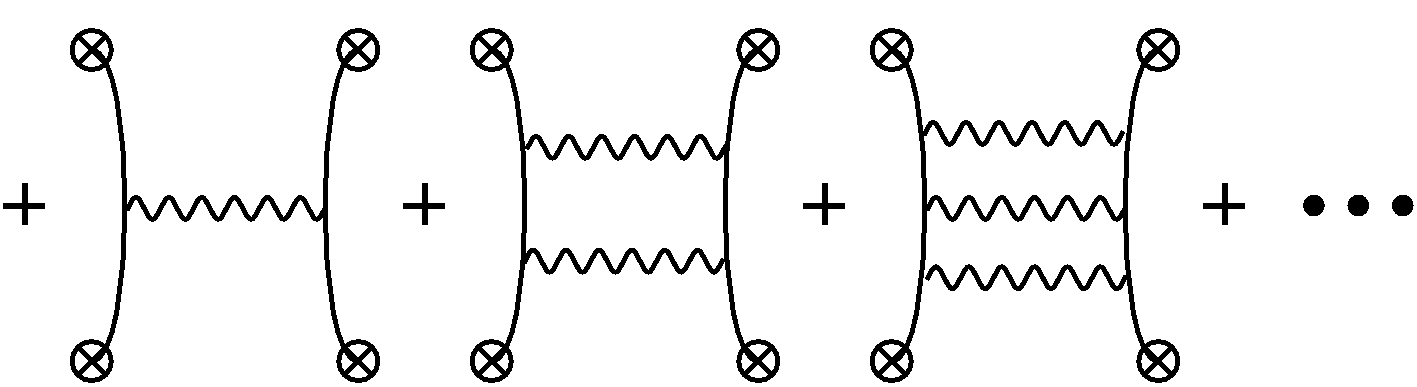}
\end{gathered} 
\end{split}
\end{equation*}
where ${\cal V}_0$ denotes the normalized {\it Virasoro identity block}. The wavy lines indicate exchanges of reparametrization modes between two pairs of external operators. At leading order in large $c$, we get
\begin{equation}
\begin{split}
 {\cal V}_0 \big{|}_{{\cal O}(1)} &= \sum_{n\geq 0} \;\big{\langle} {\cal B}_{\epsilon,h}^{(n)}(z_1,z_2) \, {\cal B}_{\epsilon,h}^{(n)}(z_3,z_4) \big{\rangle} \Big{|}_{{\cal O}\left((\frac{h^2}{c})^n\right)}   + {\cal O}\left(c^{-1}\right)\\
   &= \sum_{n\geq 0} \frac{1}{n!} \left( \frac{2h^2}{c} z^2 \; {}_2F_1(2,2,4,z) \right)+ {\cal O}\left(c^{-1}\right) \\
   &= \text{exp} \left( \frac{2h^2}{c} z^2 \; {}_2F_1(2,2,4,z) \right)+ {\cal O}\left(c^{-1}\right) \,,
\end{split}
\end{equation}
where we used \eqref{eq:Bp} to replace ${\cal B}_{\epsilon,h}^{(n)} \rightarrow \frac{1}{n!} \, \big({\cal B}_{\epsilon,h}^{(1)}\big)^n$ and included an additional factor of $n!$ to account for permutations of contractions.
This is simply the exponentiated version of the physical block in \eqref{eq:2dBSb}.

\vspace{10pt}
\section{Details on calculations in section \ref{sec:EEcorr}}

In this appendix we collect some details on calculations outlined in the main text.

\subsection{Derivation of ${\cal G}_{00,00}(P_1,P_2)$}
\label{sec:deriveG0000}

In order to prove \eqref{eq:G0000res}, we first note that the general two-point function \eqref{eq:Gcorr} can be written as 
\begin{equation}\label{eq:calc1}
\begin{split}
 & {\cal G}_{\mu\nu,\rho\sigma}(P_1,P_2) \\
 &\quad =  \frac{8 C_T}{(-2P_1 \cdot P_2)^{d+2}} \, \mathbb{P}_{\mu\nu}^{AB}(P_1)\, \mathbb{P}_{\rho\sigma}^{CD}(P_2) \Big[ \left((P_1 \cdot P_2) \eta_{BC} - P_{1,C} P_{2,B} \right) 
\left((P_1 \cdot P_2) \eta_{AD} - P_{1,D} P_{2,A} \right)  \Big] 
\end{split}
\end{equation}
To compute the time-time component of this, we note that the time-time component of the projectors takes the simple form\footnote{  This can be expressed more covariantly by noting that $m^A(\tau) = \Omega^A{}_B (P^B - Y^B)$, where the rotation matrix $\Omega_{AB} = \delta_{A}^{II} \delta_{B}^0 - \delta_{A}^{0} \delta_{B}^{II}$.
}
\begin{equation}
  \mathbb{P}_{\tau\tau}^{AB}(P) = m^A m^B- \frac{1}{d} \, \eta^{AB}  \, ,\qquad 
  \text{where } \,\, m^A \equiv m^A(\tau) = (0,0,\cos \tau, - \sin \tau, \vec{0}).
\end{equation}
Using this, we can write out the terms in \eqref{eq:calc1} explicitly. The terms simplify upon using the following identities:
\begin{equation}
\begin{split}
    \eta_{AB}\, m^A m^B &= 1 \,, \qquad \eta_{AB} \,m^A_1 m^B_2= \cos(\tau_1 - \tau_2) \,,\\
   \eta_{AB} \,m^A P^B &= 0 \,,\qquad\, \eta_{AB} \,m^A_1 P^B_2 = -\sin(\tau_1-\tau_2) \,.
\end{split}
\end{equation}
Finally, writing all $\sin^2(\tau_1 - \tau_2) = 1 - \cos^2(\tau_1-\tau_2)$, one finds
\begin{equation}
\begin{split}
  {\cal G}_{00,00}(P_1,P_2) &=  \frac{8C_T}{(-2P_1 \cdot P_2)^{d+2}}  \bigg\{ 1 + \frac{2(d-1)}{d}\, \cos(\tau_1-\tau_2) \, (Y_1 \cdot Y_2) \\ 
   &\qquad\qquad - \frac{1}{d} \left( \cos^2(\tau_1-\tau_2) + (Y_1 \cdot Y_2)^2 \right)  
   + \cos^2(\tau_1-\tau_2) \, (Y_1 \cdot Y_2)^2 \bigg\} \,.
\end{split}
\end{equation}
Clearly, all the appearances of $\cos(\tau_1-\tau_2)$ and $Y_1 \cdot Y_2$ can be obtained by differentiating the auxiliary quantity ${\cal G}_\Delta^{a,b}(P_1,P_2)$ with respect to $a$ and $b$. The result then takes the form \eqref{eq:G0000res}.

\vspace{10pt}
\subsection{Derivation of ${\cal G}_\Delta^{a,b}(\omega_E,k)$}
\label{sec:ohyaCalc}

We provide here some details on the Fourier transform leading to \eqref{eq:GDeltaMom2}. Essentially, the calculation is the same as a Euclidean version of Appendix A.2 of \cite{Ohya:2016gto} with the additional replacement of Bessel functions $I_{\omega_E} (u)\rightarrow  I_{\omega_E}(a u)$ and $K_{ik}(u)  \rightarrow b^{-(d-2)/2} K_{ik}(b u)$ to take the propagator off-shell. We now recall some of the essential steps. 

We start with the following identity that follows from orthonormality and completeness of the basis functions $f_{\omega_E,k,p_\perp}(P)$:
\begin{equation}
\label{eq:calc2}
\begin{split}
   &{\cal G}^{a,b}_\Delta(\omega_E,k) \, f_{\omega_E,k,p_\perp}(P) = \int dP' \; {\cal G}_\Delta^{a,b}(P,P') \, f_{\omega_E,k,p_\perp}(P') \\
   &\quad = \left( \frac{4k \, \sinh(\pi k)}{\pi} \right)^{1/2}  \, \int_0^{2\pi} d\tau' \int_0^\infty \frac{d\rho'}{\rho'^{d-1}} \int d^{d-2}x_\perp' \; \frac{\rho'^{\frac{d-2}{2}} \, K_{ik}(|p_\perp| \, \rho') \, e^{i(\omega_E \tau' + p_\perp \cdot \, x_\perp')}}{\left[-2a \, \cos(\tau-\tau') + b \, \frac{\rho^2+\rho'^2+|x_\perp-x_\perp'|^2}{\rho\rho'} \right]^\Delta }  \\
   &\quad =  \left( \frac{4k \, \sinh(\pi k)}{\pi} \right)^{1/2}  \, \frac{1}{2^\Delta \Gamma(\Delta)}\, e^{i(\omega_E \tau + p_\perp \cdot \, x_\perp)} \int_ 0^{2\pi} d\tau' \,\int_0^\infty dz \, z^{\Delta - 1}  e^{i\omega_E \tau'+ az\, \cos \tau } \\
   &\qquad\quad \times \int_0^\infty d\rho' \, \rho'^{-d/2} \, K_{ik}(|p_\perp| \, \rho') \, e^{-\frac{b  z}{2\rho\rho'}(\rho^2+\rho'^2) - \frac{\rho\rho'}{2bz} p_\perp^2 } \, \left( \frac{2\pi \rho \rho'}{bz} \right)^{(d-2)/2} 
 \end{split}
\end{equation}
where we introduced the Schwinger parameter $z$ to exponentiate the denominator of the two-point function in the last step. We now use the following identities:
\begin{equation}
\begin{split}
  \int_0^{2\pi} d\tau' \; e^{i\omega_E \tau'+ az \, \cos \tau'} &= 2\pi  \, I_{\omega_E}(az)\,,\\
  \int_0^\infty \frac{d\rho'}{\rho'} \, K_{ik}(|p_\perp| \, \rho') \, e^{-\frac{b  z}{2\rho\rho'}(\rho^2+\rho'^2) - \frac{\rho\rho'}{2bz} p_\perp^2 } 
  &= 2 \, K_{ik}(bz) K_{ik}(|p_\perp|\, \rho) \,.
    \end{split}
\end{equation}
Inserting these in \eqref{eq:calc2} yields a prefactor that is precisely $f_{\omega_E,k,p_\perp}(P)$. We can thus strip this factor off of \eqref{eq:calc2} and write:
\begin{equation}
\begin{split}
  {\cal G}^{a,b}_\Delta(\omega_E,k) &=\frac{(2\pi)^{d/2}}{2^{\Delta-1} \Gamma(\Delta)}\, b^{-\frac{d-2}{2}}  \, \int_0^\infty dz \, z^{\Delta - \frac{d}{2}} \, I_{\omega_E}(az) \, K_{ik}(bz) \\
  &= \frac{\pi^{d/2}}{\Gamma(\Delta)} \frac{a^{\omega_E}}{b^{\omega_E+\Delta}} \, \frac{|\Gamma(\alpha)|^2}{\Gamma(\omega_E+1)}  \; {}_2F_1\left(\alpha,\alpha^*,\omega_E+1;\frac{a^2}{b^2}\right)_\text{reg.}
 \end{split}
\end{equation}
where $\alpha \equiv \frac{1}{2} \left(\Delta- \frac{d-2}{2} + ik + \omega_E \right)$. The last integral is strictly valid only when $\text{Re}(\alpha) >  0$ and $a<b$.\footnote{  See, for example, section 6.576 on page 684 of \cite{gradshteyn2007}.} Especially when we take the limit $a,b \rightarrow 1$, we need to regularize the expression and extract the finite piece. We achieve this as follows. When computing \eqref{eq:AnyDimRes}, we first take derivatives with respect to $a$ and $b$ as instructed by \eqref{eq:G0000res}. We then replace any instances of hypergeometric functions according to
\begin{equation}
   {}_2F_1\left(\alpha,\alpha^*,\omega_E+1;\frac{a^2}{b^2}\right) \rightarrow \frac{\Gamma(\omega_E + 1)\Gamma(\omega_E + 1 - \alpha - \alpha^*)}{\Gamma(\omega_E + 1 - \alpha)\Gamma(\omega_E + 1 - \alpha^*)}
\end{equation}
It is now well-defined to set $a=b=1$ in the resulting expression.

\vspace{10pt}

\subsection{Dimensional regularization for even dimensions}
\label{sec:EvenDimReg}

In this section we describe how to perform the dimensional regularization of \eqref{eq:AnyDimRes} in the case of even dimensions. We set $d=2m - \varepsilon$ and wish to extract the finite term as $\varepsilon \rightarrow 0$. 

For the divergent $\Gamma$-function, we have $\Gamma(-\frac{d}{2}) \rightarrow \frac{(-1)^m}{m!} \left( \frac{2}{\varepsilon} + \psi(m+1) + {\cal O}(\varepsilon) \right)$. A similar expansion of the other $d$-dependent terms yields the following result for the ${\cal O}(\varepsilon^0)$ term:
\begin{equation}
\label{eq:EvenDimRes}
\begin{split}
   &{\cal G}_{00,00}(\omega_E,k)_\text{finite} \\
   &\quad=  \frac{2C_T\,(-1)^{m-1}\pi^m(m-1)}{4^m (2m+1)m! (2m-1)!} \left( \prod_{j=1}^{m-2} \left( (m+\omega_E-1-2j)^2 + k^2 \right) \right) \\
   &\qquad\quad \times \left( k^2 +m^2 \right) \left( k^2 + (m-1)^2 \right)   \bigg\{  \psi\left( \frac{\omega_E+i k +m-1}{2}\right) + \psi\left( \frac{\omega_E-i k +m-1}{2}\right)  \\
   &\qquad \qquad \qquad\qquad\qquad\qquad\qquad\quad\;\;+ \psi\left( \frac{\omega_E+i k -m+3}{2}\right) + \psi\left( \frac{\omega_E-i k -m+3}{2} \right)\bigg\}
\end{split}
\end{equation}
where $\psi(z)$ is the digamma function.
In writing this, we have dropped various contact terms (i.e., terms involving only polynomials in $\omega_E$ and $k$, but no singular functions such as $\psi(...)$).

\vspace{10pt}

\bibliographystyle{JHEP}

\providecommand{\href}[2]{#2}\begingroup\raggedright\endgroup


\end{document}